\documentclass[floats,floatfix,showpacs,amssymb,prd,twocolumn,superscriptaddress,nofootinbib,nolongbibliography,reprint]{revtex4-2}

\usepackage{amssymb,amsmath,verbatim,mathtools,needspace,enumitem,etoolbox,graphicx,physics,microtype,afterpage,xspace,tabularx,lmodern,multirow,upgreek}
\usepackage[normalem]{ulem}
\usepackage[dvipsnames, usenames]{xcolor}
\definecolor{linkcolor}{rgb}{0.0,0.3,0.5}
\usepackage[unicode, colorlinks=true, linkcolor=linkcolor, citecolor=linkcolor, filecolor=linkcolor, urlcolor=linkcolor, linktocpage, breaklinks]{hyperref}
\usepackage[all]{hypcap}
\usepackage{subfigure}
\usepackage[T1]{fontenc}
\usepackage[utf8]{inputenc}
\usepackage{diagbox} %
\usepackage{tikz}
\usepackage{soul}

\usepackage[usenames,dvipsnames]{xcolor}
\hypersetup{colorlinks=true,citecolor=romared,linkcolor=romared,urlcolor=romared}

\definecolor{romared}{RGB}{142,0,28}

\newcommand{\pd}{\partial}
\newcommand{\be}{\begin{equation}}
\newcommand{\ee}{\end{equation}}

\def\be{\begin{equation}}
\def\ee{\end{equation}}
\newcommand{\beq}{\begin{eqnarray}}
\newcommand{\eeq}{\end{eqnarray}}
\usepackage{aas_macros}
\usepackage{makecell}
\usepackage{soul}
\usepackage{amssymb}
\usepackage{lipsum}
\allowdisplaybreaks

\newcommand{\msun}{M_{\odot}}

\newcommand{\dm}{DM}
\newcommand{\edotdf}{\dot{E}_{\textrm{DF}}}
\newcommand{\edotgw}{\dot{E}_{\textrm{GW}}}
\newcommand{\edotorb}{\dot{E}_{\textrm{orbit}}}
\newcommand{\BH}{BH}
\newcommand{\bh}{black hole}

\newcolumntype{Y}{>{\centering\arraybackslash}X}

\begin{document}
\title{Extreme mass ratio inspirals in rotating dark matter spikes}

\begin{abstract}
Gravitational wave (GW) signals from extreme mass ratio inspirals (EMRIs) are a key observational target for the Laser Interferometer Space Antenna (LISA). The waveforms may be affected by the astrophysical environment surrounding the central black hole (BH), and in particular by the surrounding dark matter (DM) distribution. In this work, we consider the effect of a rotating DM ``spike'' around a central Kerr BH, and assess its detectability with LISA. Using a fully relativistic model for the rotating spike, we investigate its effect on the inspiral and hence on the emitted GW signals. We compute dephasings and mismatches to quantify how the spin of the primary BH affects the binary dynamics and the gravitational waveform. We show that the modifications due to the spin of the primary BH improve the detection prospects of DM spikes with LISA, and must be taken into account for future parameter estimation studies. We also estimate within post-Newtonian theory how the environment affects the background metric, and show that this effect is mostly negligible for the systems we consider. 
\end{abstract}
\newcommand{\jhu}{William H.\ Miller III Department of Physics and Astronomy, Johns Hopkins University, \\ 3400 N. Charles Street, Baltimore, Maryland, 21218, USA}

\author{Soumodeep Mitra}
\email{soumodeep.mitra@coyotes.usd.edu}
\affiliation{Department of Physics, University of South Dakota, Vermillion, SD 57069, USA}

\author{Nicholas Speeney}
\email{nspeene1@jhu.edu}
\address{\jhu}

\author{Sumanta Chakraborty}
\email{tpsc@iacs.res.in}
\affiliation{School of Physical Sciences, Indian Association for the Cultivation of Science, Kolkata 700032, India}

\author{Emanuele Berti}
\email{berti@jhu.edu}
\address{\jhu}

\date{\today}
\maketitle
\section{Introduction}\label{sec:intro}

The direct detection of gravitational waves (GWs) from the merger of binary black holes (BHs) by the LIGO-Virgo-KAGRA (LVK) collaboration has opened exciting experimental probes into general relativity and astrophysics~\cite{LIGOScientific:2016aoc,LIGOScientific:2020ibl,LIGOScientific:2018mvr,LIGOScientific:2021djp}. In particular, observed GW signals may offer insights into the existence and nature of exotic compact objects as BH alternatives~\cite{Cardoso:2019rvt,Liebling:2012fv,Cardoso:2016oxy,Chakravarti:2024ncc}, the spectral instability and non-linearity in the quasi-normal mode frequencies~\cite{Jaramillo:2020tuu,Cheung:2021bol,Jaramillo:2021tmt,Destounis:2021lum,Sarkar:2023rhp}, and, importantly, serve as probes of the astrophysical environments surrounding BHs~\cite{Barausse:2014tra,Speeney:2022ryg,Kavanagh:2020cfn,Speri:2022upm,Cardoso:2021wlq,Cardoso:2022whc,Chakravarti:2025xaj,Gliorio:2025cbh,Figueiredo:2023gas,Khalvati:2024tzz}. Thus far, tests of these effects have been limited by the sensitivity range of current GW detectors. Detection prospects will increase, however, as next-generation detectors such as the space based Laser Interferometer Space Antenna (LISA)~\cite{LISA:2017pwj,LISA:2022kgy,LISA:2022yao} (operating in the mHz range) and the ground based Einstein Telescope~\cite{ET:2019dnz,Punturo:2010zz,Abac:2025saz} and Cosmic Explorer~\cite{Reitze:2019iox} (operating in the $\sim 1\text{Hz}-10^3\text{Hz}$ range) are added to the network. A key observational target for LISA will be extreme mass-ratio inspiral (EMRI) systems, comprised of a small $\sim 1-100$$\msun$ body orbiting a supermassive BH with mass $\sim 10^5M_\odot-10^8M_\odot$. LISA EMRIs will undergo $\sim10^5$ orbital cycles throughout their lifetime, offering exquisitely precise tests of their orbital properties with GWs. For EMRI waveforms to match the precision required for LISA observations, much work has been dedicated to the precise modeling of EMRIs in both vacuum~\cite{Poisson:2011nh,vandeMeent:2017zgy, Barack:2018yvs,Wardell:2021fyy,LISAConsortiumWaveformWorkingGroup:2023arg,Albertini:2023aol} and non-vacuum scenarios~\cite{Cardoso:2022whc,Cardoso:2021wlq,Brito:2023pyl,Kocsis:2011dr,Duque:2023seg}. EMRIs primarily form in the central regions of galaxies, where a host of non-vacuum environments are expected to exist. In particular, the distribution of dark matter (DM) in this region may influence the secondary body, and leave observable imprints on the GW signal. 

Although the galactic scale distribution of \dm\ is well known from various simulations and observational tests~\cite{Hernquist:1990be,Navarro:1995iw,Navarro:1996gj}, the behavior of \dm\ near the galactic center is still an unresolved question. Indeed, observations of dwarf galaxies~\cite{Moore:1994yx,Flores:1994gz,Marchesini:2002vm} give rise to tensions (such as the ``core-cusp problem''~\cite{de_Blok_2009}) with the expected nature of cold DM near the galactic center, giving rise to various alternate DM models~\cite{Ferreira:2020fam, Hui:2016ltb}. At the same time, even in the cold DM regime, galactic-scale simulations do not resolve the effect of the central supermassive BH at the center, which is expected to modify the distribution of \dm\ in this region. Notably, the BH may deepen the gravitational well near the center, allowing the DM to develop an over-dense cusp, or ``spike''.

The formation and evolution of such DM spikes around massive BHs was initially studied by Gondolo and Silk~\cite{Gondolo:1999ef}, who considered the adiabatic growth of supermassive non-rotating BHs in the center of a DM halo in a Newtonian framework. They found that the DM density distribution contracts inwards, forming the ``spike'' structure, which sharply drops off at $r=8{\rm{M_{BH}}}$, where ${\rm{M_{BH}}}$ is the mass of the central BH. Their formalism was later made fully relativistic by Sadeghian et al.~\cite{Sadeghian:2013laa}, where important differences in the spike distribution were revealed. In particular, the DM spike distribution can extend down to $r=4{\rm{M_{BH}}}$, with higher densities compared to the Newtonian counterpart. Further work by Ferrer et al.~\cite{Ferrer:2017xwm} generalized the Sadeghian treatment to the Kerr case, where the centrally grown BH was allowed to spin. They found that as the BH spin increases, DM densities along the equator can grow even higher than the Schwarzschild case, and the resulting density distribution naturally develops a toroidal structure around the BH's axis of rotation. In all of these works, the primary focus was to investigate how DM annihilation signals could be enhanced by the increased DM density in the galactic center. 

Recently, much progress has been made towards understanding how these spikes and the \dm\ environment, in general, could affect the GW signal from a secondary body orbiting the massive central BH~\cite{Miller:2025yyx}. As the secondary body moves through the environment, its motion gives rise to a gravitational wake of the \dm\ particles, producing a drag force that slows down the body. This effect, known as ``dynamical friction'' (DF), was first studied by Chandrasekhar~\cite{Chandrasekhar:1943ys} and later by various authors~\cite{Kim:2007zb, Barausse:2007dy, Vicente:2022ivh,Ostriker:1998fa, Mitra:2023sny}. Alongside GW emission, the existence of such a dissipative force will act as a source of energy loss, leading to observable signatures in the GW data. The effect of the environment on GW phasing was initially investigated by Eda et al.~\cite{Eda:2013gg,Eda:2014kra} for the Gondolo-Silk spike distributions and later extensively studied in literature for the DM distribution in a static Schwarzschild spacetime~\cite{Cardoso:2021wlq, Cardoso:2022whc, Speeney:2022ryg, Speeney:2024mas, Cole:2022yzw, Cole:2022ucw, Figueiredo:2023gas, Gliorio:2025cbh, Chakraborty:2024gcr,Hu:2023oiu,Rahman:2023sof}. Additionally, other authors investigated the dynamic evolution of the DM spike as the secondary body inspirals~\cite{Coogan:2021uqv,Kavanagh:2020cfn,Bertone:2024wbn}, and how this might affect the GW signal. All of these studies, however, were limited to the case of a non-rotating BH in a spherically symmetric DM environment.

In astrophysical scenarios, on the other hand, we expect the central \bh\ and \dm\ halo to be rotating. Observationally, the supermassive BHs at the center of active galactic nuclei have been found to be rapidly rotating~\cite{Jones:2020nnx, Reynolds:2013qqa, Done:2013pha, Reynolds:2019uxi}, and GW measurements from the LVK collaboration have revealed the population of stellar mass BHs to also be spinning~\cite{LIGOScientific:2020kqk, Biscoveanu:2020are}.For generically spinning EMRIs surrounded by matter, the central BH's rotation will also induce rotation in surrounding matter distributions through frame dragging effects~\cite{Sadeghian:2013laa,Ferrer:2017xwm}. This change in the environment further induces changes in the secondary body's inspiral, and hence affects the GW signals that LISA will detect. Therefore, modeling both the background spacetime and the environment in the presence of rotation is important, since any prediction of GW signals from EMRIs in the presence of these environments will be impacted by the effects of rotation.

There have so far been only a few attempts to include rotational effects into the environment, and in particular to investigate the implications for GWs~\cite{Dyson:2025dlj,Khalvati:2024tzz}. To address this gap, in this work we consider the impact of rotating DM spikes on EMRIs with a spinning central BH, and investigate how GW signals are modified in response to the environment. While a fully self-consistent treatment of rotating environments is not yet available, we include as much relativistic modeling as possible in two key ways: (i) by employing the relativistic formalism in~\cite{Ferrer:2017xwm} to compute the rotating DM spike distribution, assuming an adiabatic growth model; (ii) by using the \texttt{FastEMRIWaveform (FEW)} package~\cite{Chua:2018woh,Chua:2020stf,Katz:2021yft,michael_l_katz_2023_8190418} to generate fully relativistic GW signals in a rotating background, including the added environmental effect due to DF. 

By computing GW dephasing with the rotating DM spike, we demonstrate that adding rotation enhances observable properties of the binary. In particular, higher spike densities lead to a larger DF force, having significant impact on the EMRI's observation time in the LISA band. Further, rotating spikes increase GW dephasing compared to their non-rotating counterparts, thus increasing detection prospects. To examine the observability of DM induced effects with LISA, we also perform mismatch computations for both prograde and retrograde orbits. We find again that rotation enhances LISA's chance of observing DM induced effects, and demonstrate that rotating EMRIs allow us to observe more diffuse DM halos compared to their non-rotating counterparts. Our results are further corroborated using the growth of the signal-to-noise ratio (SNR) with rotation and mismatch between gravitational waveforms with and without rotation.

The paper is organized as follows: we start in Section \ref{DMKerr} by providing an overview of the DM spike growth in the Kerr BH background. In Section \ref{sec:GW_treatment}, we detail the key effects that the DM spike has on the EMRI's orbital dynamics, and include these effects in the gravitational waveform using the \texttt{FEW} package. Subsequently, in Section \ref{sec:PN_estimate}, we estimate the modifications to the background metric due to the DM spike profile, and show that such changes to the background geometry are negligible for our purposes. Finally, in Section \ref{sec:results} we use our environmental gravitational waveform model to investigate the detectability and observability of DM spike distributions with LISA via mismatch and SNR computations. We conclude with a discussion of possible directions for future work. 
Throughout this paper, we will use a mostly positive $(-,+,+,+)$ metric signature, and define units such that $G=c=1$, unless otherwise specified. Greek indices span all four spacetime coordinates, while Latin indices run over the spatial coordinates.

\section{Dark Matter Spike profile in the Kerr background}\label{DMKerr}

In this section, we present the essential elements in the computation of the relativistic DM density profile in the presence of a rotating Kerr BH background; for more detail, we refer the reader to~\cite{Ferrer:2017xwm}. 
We start with an initial DM distribution around the supermassive BH, which we take to be the Hernquist profile~\cite{Hernquist:1990be}. It is given by
\begin{equation}
\label{eq:hernquist_profile}
\rho_{\rm ini}(r)=\frac{({\rm{M_{halo}}}/2\pi {\rm{r_s}}^3)}{(r/{\rm{r_s}})(1+r/{\rm{r_s}})^3}\,,
\end{equation}
where ${\rm{M_{halo}}}$ is the total mass of the DM halo, and ${\rm{r_s}}$ is a scale radius of the DM distribution. We could in principle choose other more widely used initial density profiles such as the Navarro-Frenk-White (NFW) profile~\cite{Navarro:1995iw, Navarro:1996gj}, or a power law profile. We restrict our analysis to the Hernquist profile, however, since it mimics the NFW behavior near the central region, and has the added advantage of a closed-form expression for the phase space distribution function shown in Eqs.~\eqref{eq:Hern_distribution_function1} and \eqref{eq:Hern_distribution_function2}. This choice is also sufficient to highlight key features of the DM spike growth near a rotating BH, which we show below.

Initially, the DM density profile is spherically symmetric, such that the action variables associated with a DM particle having energy per unit mass $E_{\rm h}$, angular momentum per unit mass $L_{\rm h}$ and $z$-component of the angular momentum per unit mass $L_{\rm z,h}$ are given as 
\begin{equation}
\begin{aligned}
\label{eq:halo_actions}
I_r^{\rm h}&=\oint \text{d}r \sqrt{2E_{\rm h}-2\Phi-\frac{L_{\rm h}^2}{r^2}}\,,
\\
I_\theta^{\rm h}&=\oint \text{d}\theta \sqrt{L_{\rm h}^2-\frac{L_{\rm z,h}^2}{\sin^2{\theta}}}=2\pi(L_{\rm h}-|L_{\rm z,h}|)\,,
\\
I_\phi^{\rm h}&=\oint \text{d}\phi L_{\rm z,h}=2\pi L_{\rm z,h}\,,
\end{aligned}
\end{equation}
where $\Phi$ is the Newtonian gravitational potential of the DM distribution. The above action integrals are adiabatically invariant quantities, so we may equate these with the action integrals \textit{after} the growth of a central BH within the DM distribution. For bound orbits, after the adiabatic growth, we have the following action integrals: 
\begin{equation}
\begin{aligned}
\label{eq:Kerr_actions}
I^{\rm K}_r&
=\oint \text{d}r \frac{\sqrt{V(r)}}{\left(1+\frac{a^2}{r^2}-\frac{2{\rm{M_{BH}}}}{r}\right)}\,,
\\
I^{\rm K}_\theta&
=\oint \text{d}\theta \sqrt{U(\theta)}\,,
\\
I^{\rm K}_\phi&
=2\pi L_z\,.
\end{aligned}
\end{equation}
Here ${\rm{M_{BH}}}$ is the central BH's mass and $a$ is the Kerr parameter, related to the BH's angular momentum $J$ via $a=J/{\rm{M_{BH}}}$. The effective radial and angular potentials for an orbiting particle in the Kerr geometry, $V(r)$ and $U(\theta)$, can be written in terms of the orbital quantities ($\mathcal{E}$, $C$, $L_z$), the energy per unit mass, Carter constant per unit $(\text{mass})^2$, and angular momentum along the $z$-direction per unit mass, respectively:
\begin{equation}
\begin{aligned}
\label{eq:effective_potentials}
V(r)&=\left( 1+\frac{a^2}{r^2}+\frac{2 {\rm{M_{BH}}}a^2}{r^3}\right)\mathcal{E}^2-\frac{\Delta}{r^2}\left(1+\frac{C}{r^2} \right)\\
&~~~ +\frac{a^2 L_z^2}{r^4}-\frac{4 {\rm{M_{BH}}}a \mathcal{E}L_z}{r^3}\,,
\\
U(\theta)&=C-\frac{L_z^2}{\sin^2\theta}-a^2(1-\mathcal{E}^2)\cos^2\theta\,,
\end{aligned}
\end{equation}
with $\Delta=r^2-2{{\rm{M_{BH}}}}r+a^2$. After the adiabatic growth of the central BH, particles in the final DM distribution have energies and angular momenta which have been modified in response to the BH. Hence, we wish to find the energy $E_{\rm h}$ and angular momenta $L_{\rm h},L_{\rm z,h}$ within the DM halo as a function of the Kerr orbital parameters $(\mathcal{E},C,L_z)$. First, equating $I^{\rm h}_\phi=I^{\rm K}_\phi$, and using the expressions in Eqs.~\eqref{eq:halo_actions} and \eqref{eq:Kerr_actions}, we find $L_z=L_{\rm z,h}$.
Using this relation, along with $I^{\rm h}_{\theta}=I^{\rm K}_{\theta}$, we can determine the angular momentum in the halo $L_{\rm h}$ in terms of the Kerr orbital parameters as 
\begin{equation}
\label{eq:L_relation}
L_{\rm h}=|L_{\rm z,h}|+\frac{I^{\rm K}_\theta(\mathcal{E},C,L_z)}{2\pi}\,.
\end{equation}
Finally, using this value of $L_{\rm h}$, and equating $I^{\rm h}_{r}$ with $I^{\rm K}_{r}$ from Eqs.~\eqref{eq:halo_actions} and \eqref{eq:Kerr_actions}, we obtain $E_{\rm h}$ as a function of the Kerr parameters. The radial integrals in this case cannot be obtained analytically, so we numerically compute them and use a root-finding algorithm to determine $E_{\rm h}(\mathcal{E},C,L_z)$. 

\begin{figure}[t]
\label{fig:DM_spikes_Milky_Way}
  \centering

\includegraphics[scale=0.491]{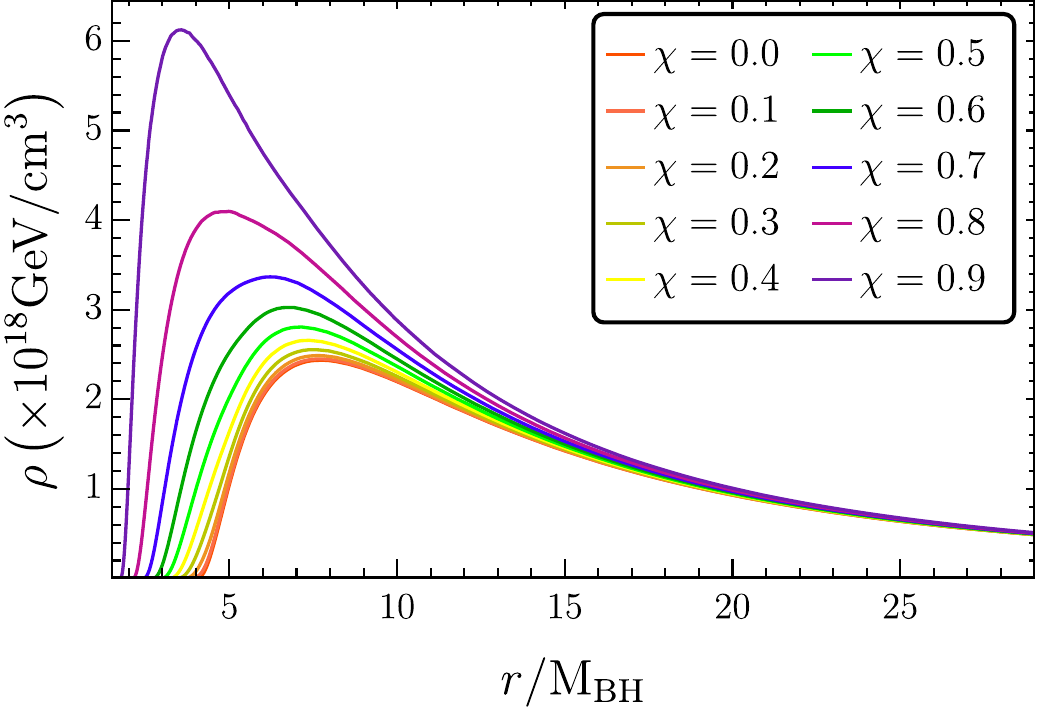}
  
  \caption{Density distribution of DM along the equatorial plane for a ``Milky-Way like'' DM halo, with ${\rm{M_{halo}}}=10^{12}M_\odot$, ${\rm{r_s}}=20\text{kpc}$, and ${\rm{M_{BH}}}=4\times10^6 M_{\odot}$. As the dimensionless spin $\chi$ increases, the DM density both increases and moves inwards toward the innermost marginally bound Kerr orbit, $(r_{\rm mb}/{\rm{M_{BH}}})=2-\chi+2\sqrt{1-\chi}$.}
\end{figure}

Subsequently, from the non-relativistic distribution function for the initial Hernquist density profile, one can obtain the relativistic DM density profile after the central BH grows. The initial non-relativistic distribution function associated with the Hernquist density profile is given by  
\begin{equation}
\label{eq:Hern_distribution_function1}
f_{\rm ini}(\varepsilon)=\frac{{\rm{M_{halo}}}F(\varepsilon)}{\sqrt{2}(2\pi)^3({\rm{M_{halo}}}{\rm{r_s}})^{3/2}},
\end{equation}
with the function $F(\varepsilon)$ being
\begin{equation}
\label{eq:Hern_distribution_function2}
F(\varepsilon)=\frac{\sqrt{\varepsilon}}{(1-\varepsilon)^2}\left[(1-2\varepsilon)(8\varepsilon^2-2\varepsilon-3)+\frac{3\sin^{-1}\sqrt{\varepsilon}}{\sqrt{\varepsilon(1-\varepsilon)}}\right]\,.
\end{equation}
The variable $\varepsilon$ is a dimensionless relative energy, related to the energy $E_{\rm h}$ (per unit mass) of DM particles by $\varepsilon={\rm{r_s}}(1-E_{\rm h})/{\rm{M_{halo}}}$. Adiabatic growth preserves the functional form of the distribution function, so we may use the previous relation connecting the energy of a particle in the DM halo with the Kerr orbital parameters, i.e., $E_{\rm h}=E_{\rm h}(\mathcal{E},C,L_{z})$, to obtain the relativistic distribution function $f(\mathcal{E},C,L_{z})$. Given this, we can obtain the two non-trivial components of the mass current density, $J_{\mu}$, as
\begin{equation}
\begin{aligned}
\label{eq:mass_current}
J_t(r,\theta)&=\frac{-2}{r^2\sin \theta}\int \text{d}\mathcal{E} \text{d}C\text{d}L_z \frac{\mathcal{E}f(\mathcal{E},C,L_z)}{\sqrt{V(r)}\sqrt{U(\theta)}}\,,
\\
J_\phi(r,\theta)&=\frac{2}{r^2\sin \theta}\int \text{d}\mathcal{E} \text{d}C\text{d}L_z \frac{L_z f(\mathcal{E},C,L_z)}{\sqrt{V(r)}\sqrt{U(\theta)}}\,.
\end{aligned}
\end{equation}
For chosen values of $(r,\theta)$, we evaluate Eqs.~\eqref{eq:mass_current} and take $\rho=\sqrt{-g^{\mu \nu}J_{\mu}J_{\nu}}$ to obtain the final DM distribution surrounding the Kerr BH. 

The resulting density profiles for various choices of the dimensionless spin parameter $\chi\equiv (a/{\rm{M_{BH}}})$ are shown in Fig.~\ref{fig:DM_spikes_Milky_Way} for a DM halo with ``Milky-Way like'' parameters, ${\rm{M_{halo}}}=10^{12} M_\odot$, ${\rm{r_s}}=20\text{kpc}$, and ${\rm{M_{BH}}}=4\times10^6 M_\odot$. Along the equatorial plane the density takes on its highest values, with enhancement of the spike occurring as the BH spin is increased. Additionally, the spin allows the spike to get closer to the central BH, and imbues the density distribution with a toroidal structure~\cite{Ferrer:2017xwm}.

To study the effect of the spike profile on GW astrophysics, we fit the numerically computed equatorial density profiles for various configurations of the halo mass, scale radius, BH mass, and BH spin. We first find scaling relations for the halo parameters and central BH mass by generating many DM spikes for various parameter configurations. We then use this catalog of spikes to obtain simple scaling relations by comparing how the density profiles change as each parameter is individually varied, as in Ref.~\cite{Speeney:2022ryg}. These scalings are given as
\begin{equation}
\label{eq:spike_fit}
\rho(r)=\left(\frac{{\rm{M_{halo}}}}{10^{12}M_\odot}\right)^{\frac{1}{3}}\left(\frac{{\rm{M_{BH}}}}{4\times 10^6M_\odot}\right)^{-\frac{5}{3}}\left(\frac{{\rm{r_s}}}{20\text{kpc}}\right)^{-\frac{2}{3}}\bar{\rho}(r)~,
\end{equation}
where $\bar{\rho}$ is found by fitting a reference curve with the same ``Milky-Way like'' parameters as in Fig. \ref{fig:DM_spikes_Milky_Way}. 
The resulting fitting function for $\bar{\rho}(r)$ is given by the following expression: 
\begin{equation}
\label{eq:DMspike_fitting_function}
\bar{\rho}(r)=H(r-r_{\rm mb})\left(\sum_{i=0}^{n}A_i(r-r_{\rm pk})^i\right)\left(\sum_{j=0}^{m}\frac{B_{j}}{r^{j}}\right)~,
\end{equation}
where $H(x)$ is the Heaviside step function, $n,m$ are chosen by hand to minimize the fitting error for each curve, $r_{\rm pk}$ is the peak of the numerically generated density curve, $(r_{\rm mb}/{\rm{M_{BH}}})=2-\chi+2\sqrt{1-\chi}$ is the location of the innermost marginally bound orbit on the equator, and the coefficients $A_i,B_i$ are determined using standard fitting packages in \texttt{Mathematica}.In all cases, we ensure the fitting error is below $0.1\%$ in the region relevant for EMRI inspirals, and we have checked that the scalings appearing in Eq. \eqref{eq:spike_fit} are robust for the range of parameters we consider. A fit in the form of Eq.~\eqref{eq:DMspike_fitting_function} is found for each choice of the dimensionless spin value $\chi$, and in the subsequent discussions, we utilize the above fits for DM spike distributions rather than interpolated data sets. 

\section{Effects of rotating Dark Matter spike on gravitational waves}
\label{sec:GW_treatment}

In this section, we focus on the rotating \dm\ spike's effect on the binary dynamics during the inspiral. Following similar treatments in~\cite{Eda:2013gg, Eda:2014kra, Speeney:2022ryg}, we first determine how the DM spikes affect orbital properties of the binary, and contribute to excess energy loss. We then incorporate these effects into the \texttt{FEW} package from the \texttt{Black Hole Perturbation Toolkit}~\cite{BHPToolkit} to generate inspiral trajectories and gravitational waveforms. Finally, we use the gravitational waveforms with the rotating DM spike distributions to derive the change in the number of cycles or ``dephasing'' in the GW signal, compute mismatches between the vacuum and non-vacuum inspirals, and show how the SNR changes in response to added rotational effects. 

The presence of a DM environment around a central Kerr BH will affect the inspiral of a secondary object around it, primarily through the DF~force. We start by considering circular orbits, since it has been shown that DF tends to circularize eccentric orbits~\cite{Becker:2021ivq} (but see also~\cite{Yue:2019ozq}, reporting an increase in eccentricity in the presence of DF). The non-zero DF force due to the environment will lead to extra energy loss in the system, in addition to the energy loss through GW emission. In the adiabatic approximation one can write down the energy balance equation for an EMRI, losing energy through GW emission and DF, as
\begin{equation}
\label{energy balance}
\edotorb = \edotgw + \edotdf\,.
\end{equation}
Here $\edotorb$ is the loss in the orbital energy of the binary, $\edotgw$ is the energy loss through GW emission, and $\edotdf$ is the energy loss due to DF. The energy loss due to DF can be written as $\edotdf = \textbf{v}\cdot\textbf{F}_{\rm{DF}}$, where $\textbf{v}$ is the velocity of the secondary, and $\textbf{F}_{\rm{DF}}$ is the DF force. In the reference frame of an asymptotic observer, the expression for the force is given by~\cite{Chandrasekhar:1943ys, Vicente:2022ivh, Speeney:2022ryg},
\begin{equation}
\label{force_DF}
{\rm{\textbf{F}_{DF}}}=-4\pi\frac{\rho(r)\eta^{2}\textbf{v}}{v^{3}}\gamma^{2}(1 + v^2)^{2}\,{\rm{ln}}\Lambda\,, 
\end{equation}
where we have taken the DF to be parallel to the velocity, typical for any frictional force. Recently, Refs.~\cite{Wang:2024cej, Dyson:2024qrq} have worked out a general scenario where the velocity is not aligned with the DF force, giving rise to a gravitational Magnus effect which affects the binary's plane of orbit. We have kept such effects aside for future study. In the above expression for the DF force, we have introduced the reduced mass $\eta\equiv\{\mu {\rm{M_{BH}}}/({\rm{M_{BH}}}+\mu)\}$, with $\mu$ being the mass of the secondary. Moreover, $\gamma=(1/\sqrt{1- v^2})$ is the Lorentz factor and the quantity $\Lambda$ is related to the maximum impact parameter $b_{\rm{max}}$ as $\ln{\Lambda} \approx b_{\text{max}}v_{\rm typ}^2/( \mu)$ with $v_{\rm typ}$ being the ``typical'' velocity of the secondary BH, such that $\rm{ln} \Lambda \sim 3$~\cite{Chandrasekhar:1943ys}.
Finally, $\rho(r)$ represents the DM spike profile, and is given in Eq.~\eqref{eq:spike_fit}.

The dominant radiative contribution during the binary inspiral is given by the radiative energy loss through GW emission, $\edotgw$. At the leading post-Newtonian (PN) order, $\edotgw$ is given by the quadrupole formula,
\begin{equation}
\label{0PN}
\edotgw^{(0)}=-\frac{32}{5}\eta^{2} x^{10}\,,
\end{equation}
where $x\equiv({\rm{M_{BH}}}\Omega_{\rm orb})^{1/3}$, with $\Omega_{\rm orb}$ being the orbital frequency, and taking the secondary mass to be negligible compared to the central supermassive BH. One can also consider higher-order PN terms~\cite{Sasaki:2003xr,Forseth:2015oua} to compare with the effect of DF, which typically appears at negative PN order, and hence becomes more and more dominant in the early inspiral. Since EMRIs begin their inspiral at significant distance and emit much of the detectable radiation close to the ISCO ($\sim 6\rm{M}_{\rm BH}$), the DF is very important at the initial stages of inspiral, while the higher PN terms are dominant in the late inspiral. While the PN expansion allows us to more easily compare the GW emission to the effect of DF order by order and can offer insight into the relative importance of DF vs higher order GW effects, it is not suitable for complete EMRI evolutions. EMRIs spend tens of thousands to millions of orbits in the LISA band, and many of these orbits are in the strong-gravity region of the central object, where the PN expansion breaks down. To achieve the necessary precision, one should calculate the GW fluxes by solving the Teukolsky equation with appropriate source terms instead. To more accurately evolve the EMRI systems we consider, we utilize fully relativistic GW fluxes by numerically solving the Teukolsky equation for various spins using the \texttt{Teukolsky} package~\cite{wardell_2025_14788956} in the \texttt{Black Hole Perturbation Toolkit}~\cite{BHPToolkit}, considering the background to be given by the Kerr metric. We incorporate the resulting GW flux into the \texttt{Trajectorybase} module of \texttt{FEW} to calculate inspiral trajectories, and subsequently use it in the generation of the gravitational waveforms. While this work was in its final stages, an updated version of \texttt{FEW} with Kerr adiabatic fluxes was released. One can also use these updated modules instead of solving the Teukolsky equation externally, as we have done. We have checked that the results presented here match those derived from the updated Kerr adiabatic flux modules of \texttt{FEW}.

Along with the energy fluxes, it is also necessary to compute the loss of angular momentum due to radiative effects. As DF introduces an additional force on the secondary, this also leads to an extra torque on the system, resulting in a loss of angular momentum given by
\begin{equation}
\label{torque}
\dot{\textbf{L}}_{\rm{DF}} = \textbf{r} \times \textbf{F}_{\rm{DF}}\,.
\end{equation}
If we consider an equatorial orbit and note that the force due to DF is opposite to the velocity, one can see that the direction of the torque is opposite to the orbital angular momentum. In a cylindrical coordinate system, with the central \BH\ spinning along the $z$ direction, the orbital angular momentum is directed along the $\pm z$ direction, for prograde and retrograde orbits respectively. One can show that the torques due to DF and GW emission both act along the $\mp z$ direction so that, in analogy with the energy balance equation, we obtain
\begin{equation}
\label{ang_mom balance}
\dot{L}_{\rm{orbit}} = \dot{L}_{\rm{GW}} + \dot{L}_{\rm{DF}}\,.
\end{equation}
Here we have dropped the vector notation from the terms, and we note that both $\dot{L}_{\rm{GW}}$ and $\dot{L}_{\rm{DF}}$ are negative quantities. The change in the orbital angular momentum can be derived from the angular momentum expression for bound orbits given in~\cite{Chandrasekhar:1985kt,Glampedakis:2002ya}.

Using Eqs.~\eqref{energy balance} and \eqref{ang_mom balance}, one can obtain the rate of change of the dimensionless semi-latus rectum $p$ and eccentricity $e$ for slightly eccentric orbits~\cite{Kennefick:1998ab}. These first order differential equations for $p$ and $e$ are obtained by inverting the expressions for $\edotorb$ and $\dot{L}_{\rm{orbit}}$, respectively, as both $E_{\rm orb}$ and $L_{\rm orb}$ depend on these two physical quantities. These equations can then be numerically solved to generate the inspiral trajectory with appropriate boundary conditions, such that the final orbit corresponds to the innermost stable circular orbit (ISCO).
We achieve this using the \texttt{Trajectorybase} module of \texttt{FEW}, with modified differential equations which incorporate the DF effect. 

Finally, we would like to point out that obtaining GW fluxes by solving the Teukolsky equation in the Kerr background essentially means that we are considering the effect of \dm\ spikes on the background geometry in the ``probe limit'', i.e., we are neglecting the effect of the \dm\ densities on the background metric itself. It is then sensible to ask about the regime of validity of our study, as in the presence of the \dm\ spike, one may expect that the background geometry is no longer given by the Kerr metric. To understand the validity of the approximation, we have studied the effect of the \dm\ spikes on the background Kerr geometry in the PN approximation. 
As we will demonstrate in the next section, for primary BHs with mass of $\gtrsim 10^{7}\msun$, the dephasing induced as a result of deviation from the ``probe-limit'' approximation, i.e. due to the change in the metric as a result of external matter, becomes comparable to the dephasing due to DF effect. Therefore the ``probe-limit'' assumption gives an upper bound on the allowed range of the primary BH mass, which we have taken to be $\lesssim 10^{6}\msun$.

On the other hand, there also exists a lower bound on the primary BH mass $\gtrsim 10^{4}\msun$, related to the ``halo feedback'' mechanism of the \dm\ environment. In this work, we have considered \textit{stationary} spike profiles corresponding to various spins of the primary BH, which the secondary inspirals through. However, the secondary itself can perturb the spike, dumping energy into the \dm\ environment due to DF. The spike profile in general also evolves as the inspiral progresses, and should be simultaneously accounted for while evolving the binary. This effect is more pronounced for systems with primary mass of $\lesssim 10^4 \msun$~\cite{Coogan:2021uqv,Kavanagh:2020cfn}. In our study, we have not considered the ``halo feedback'' of the spikes, and therefore, we have limited ourselves to systems with primary BHs of mass range $10^{5}-10^{6} \msun$ for the various dephasing and mismatch analyses. 

To investigate the detection prospects of DM induced effects in rotating systems with GWs, we perform dephasing and mismatch computations using inspiral trajectories and gravitational waveforms with and without DM spikes. We first consider the dephasing, which requires defining the number of GW cycles ($\mathcal{N}_{\rm cycle}$) during inspiral:
\begin{equation}
\label{numbercycle}
\mathcal{N}_{\rm cycle}(f)=\int_{f}^{f_{\rm ISCO}}dF\frac{F}{\dot{F}}\,.
\end{equation}
Here $F$ is an integration variable for the GW frequency, and $\dot{F}$ %
is determined from Eqs.~\eqref{energy balance} and \eqref{ang_mom balance}. Extra effects, e.g. the change in the background geometry, environments, etc., induce a quantifiable change in the number of GW cycles compared to the vacuum GR result. As a rule of thumb, any effect causing a dephasing of $\Delta \mathcal{N}_{\rm cycle}>1$ may be considered ``detectable'' in next-generation GW detectors. We will hence use the change in the number of cycles to quantify the effect of DM spikes on the gravitational waveform, in rotating and non-rotating systems.

As already emphasized, dephasing studies may tell us about the detectability of particular effects, but cannot be used as a measure of observability in GW detectors. To understand the strength of the actual signal observed in the detectors, we use SNR and mismatch computations to make more concrete statements about GW observability with LISA. The mismatch between two gravitational waveforms $h_{1}$ and $h_{2}$ is given by the expression
\begin{equation}
\label{mismatch_defn}
\mathcal{M}=1-\frac{(h_{1}|h_{2} )}{\sqrt{(h_{1}|h_{1})(h_{2}|h_{2})}}~,
\end{equation}
where $(\cdot|\cdot)$ is the inner product between real-valued time series, written as
\begin{equation}
\label{eq:inner_product}
\left(h_1|h_2\right)=4\text{Re}\int_0^\infty\frac{\hat{h}_1(f)\hat{h}^*_2(f)}{S_n(f)}\dd f.
\end{equation}
Here $S_n(f)$ is the static power spectral density (PSD) of the LISA detector~\cite{Robson:2018ifk}, and $\hat{h}$ is the Fourier transform of $h(t)$.
Similarly, one defines the SNR as
\begin{equation}
\label{SNR_defn}
\textrm{SNR} = \sqrt{(h|h)}\,. 
\end{equation}
A waveform is considered detectable if its $\text{SNR}>20$, while two waveforms are usually considered distinguishable when their mismatch is $>0.03$ .

We would like to mention here that, while our inspiral generation uses the fully relativistic Teukolsky fluxes, we have used the GW amplitude given by \texttt{AAK}, which uses a 5PN expansion of the flux. While this does not change our conclusions based on the dephasing, there may be some small quantitative changes from using relativistic waveform generation (along with relativistic inspiral) in our mismatch and SNR results. We do not expect any qualitative difference regarding our conclusions on detectability prospect, but we plan to investigate the effect of fully relativistic waveform generation in future work.

\section{Post-Newtonian estimate of the change in the metric}
\label{sec:PN_estimate}

In general, the presence of matter induces a change in the background metric which the secondary body moves through. As mentioned in the last section, \texttt{FEW} does not account for such effects when computing relativistic trajectories, so we wish to estimate when the change in the background metric becomes safe to ignore. To do so, we perform a PN computation to obtain an approximate modified ``Kerr+environment'' metric, and estimate the impact of this change on the dephasing results. Here we present key elements of the computation, and refer the reader to the Appendices for more details. 

The strategy of this section will be as follows. We first express the Kerr metric in harmonic coordinates and expand it up to 1PN order. At that same order we also quantify the modifications to the metric arising from the DM environment. The modified Kerr+environment metric leads to revised geodesic equations, which we then use to compute the dephasing for various parameter choices. We isolate the dephasing's contribution from DF and from the metric change, and with these results obtain an acceptable parameter space where the modifications to the metric due to the DM are negligible.
 
We start by taking a generic asymptotically flat metric up to 1PN order in harmonic coordinates, given by~\cite{Andersson:2022cax}
\begin{equation}
\begin{aligned}
\label{eq:PN_metric_main_text}
g^{\rm H}_{tt}&=-1+\frac{2U}{c^2}+\frac{2}{c^4}(\Psi-U^2)+\mathcal{O}\left(c^{-5}\right)\,,\\
g^{\rm H}_{tj}&=-\frac{4U_j}{c^3}+\mathcal{O}\left(c^{-5}\right)\,,\\
g^{\rm H}_{ij}&=\delta_{ij}\left(1+\frac{2U}{c^2}\right)+\mathcal{O}\left(c^{-4}\right)\,.
\end{aligned}
\end{equation}
Here and below, we will denote vector and tensor quantities expressed in the harmonic coordinates $(t_{\rm H},r_{\rm H},\theta_{\rm H},\phi_{\rm H} )$ with a subscript (or superscript) ``$\rm{H}$'', and momentarily restore factors of $c$ to keep track of the PN order. Explicit relations between the harmonic and the Boyer-Lindquist coordinates may be found in Ref.~\cite{Hergt:2007ha}. We first note that the potential $\Psi$ may be written as $\Psi=\psi+\tfrac{1}{2}\pd^{2}_{t_{\rm H}}X$,
where $X$ is a superpotential related to $U$ via the expression $\nabla^2X=U$, and $t_{\rm H}$ is the harmonic time coordinate. Since $X$ only affects $\Psi$ through a time derivative, we may ignore it, since our configuration is assumed to be stationary. To describe the DM distribution, we take its energy-momentum tensor $T_{\mu \nu}$ to be that of a perfect fluid:
\begin{equation}
\label{eq:Tmunu_perfect_fluid_main_text}
T^{\rm H}_{\mu \nu}=\frac{1}{c^2}\left(\mathcal{E}+P\right)u^{\rm H}_\mu u^{\rm H}_\nu+Pg^{\rm H}_{\mu \nu}\,,
\end{equation}
where $\mathcal{E}$ is the total energy of the fluid, $P$ the pressure, and $u^{\rm H}_\mu$ the fluid's four-velocity. We may rewrite $\mathcal{E}$ in terms of the fluid's density and specific internal energy, $\Pi$, as $\mathcal{E}=\rho\left(1+\Pi/c^2\right)$. Plugging the metric in Eq.~\eqref{eq:PN_metric_main_text} and the stress-energy tensor in Eq.~\eqref{eq:Tmunu_perfect_fluid_main_text} into the Einstein equations and expanding appropriately to 1PN order, we find that the metric potentials $U,U_i,\Psi=\psi$ obey a set of Poisson equations
\begin{equation}
\label{eq:PN_Poisson_eqs_main_text}
\begin{aligned}
\nabla^2U&=-4\pi \rho^*\,,
\\
\nabla^2 U_i&=-4\pi \rho^* v_i\,,
\\
\nabla^2\psi&=-4\pi \rho^*\left(\frac{3}{2}v^2-U+\Pi+\frac{3P}{\rho} \right)\,,
\end{aligned}
\end{equation}
where $\rho^*=\sqrt{-g_{\rm H}}\,\gamma \rho$ is a rescaled density function, $g_{\rm H}$ is the determinant of the PN metric \eqref{eq:PN_metric_main_text}, $v_i$ is the 3-velocity of the fluid (related to $u_i$ via the expression $u^i=\gamma v^i$), and $\gamma=u^t/c\simeq \left\{1+v^2/2c^2+U/c^2\right\}$.

The DM spike solutions are pressureless dust configurations, so we may set $P=0$. The solutions are also computed by assuming that the DM particles move along Kerr geodesics~\cite{Sadeghian:2013laa,Ferrer:2017xwm}. In other words, the DM is diffuse enough to have the Kerr background dominate, so we assume that the DM's impact on the metric happens at a sub-leading order. Given this, we write the Kerr+environment metric as $g_{\mu\nu}=g_{\mu\nu}^{\rm K}+\epsilon g_{\mu\nu}^{\rm env}$, where $g_{\mu\nu}^{\rm env}$ is the environmentally induced change in the metric, $g_{\mu\nu}^{\rm K}$ is the Kerr background metric, and $\epsilon$ is a bookkeeping parameter. We subsequently write the potentials appearing in Eq.~\eqref{eq:PN_metric_main_text} as $U=U^{\rm K}+\epsilon U^{\rm env}$, with similar definitions for $U_i$ and $\psi$. 
The source terms appearing in Eq.~\eqref{eq:PN_Poisson_eqs_main_text} are all $\propto \rho$, and hence are $\mathcal{O}(\epsilon)$. Given this, we find that the Kerr potentials $U^{\rm K},U_\phi^{\rm K},\psi^{\rm K}$ all satisfy Poisson equations at $\mathcal{O}(\epsilon^0,1\text{PN})$, sourced by point-mass distributions. We forego solving the $\mathcal{O}(\epsilon^0,1\text{PN})$ equations by simply matching to the expanded Kerr metric in Ref.~\cite{Hergt:2007ha}, which yields (to 1PN order in the harmonic coordinates)
\begin{equation}
\label{eq:Kerr_potentials_main_text}
U^{\rm K}=\frac{{\rm{M_{BH}}}}{r_{\rm H}},~~~U_{\phi}^{\rm K}=\frac{\chi {\rm{M_{BH}}}^2\sin^2\theta_{\rm H}}{2r_{\rm H}}\,,
\end{equation}
with $U^{\rm K}_{r}=U^{\rm K}_{\theta}=\psi^{\rm K}=0$. At $\mathcal{O}(\epsilon^1,1\text{PN})$, we find that the environmental potentials satisfy the following relations:
\begin{equation}
\label{eq:PN_Poisson_eqs_main_text1}
\begin{aligned}
\nabla^2U^{\rm env}&=-4\pi \rho\left( 1+\delta_{ij}\frac{J_{\rm H}^iJ_{\rm H}^j}{2\rho^2 c^2}+\frac{3 {\rm{M_{BH}}}}{r_{\rm H} c^2}\right)\,,
\\
\nabla^2 U_i^{\rm env}&=-4\pi  J^{\rm H}_i\left(1+\frac{2{\rm{M_{BH}}}}{r_{\rm H}c^2} \right)\,,
\\
\nabla^2\psi^{\rm env}&=-4\pi \rho\left(\frac{3}{2}\delta_{ij}\frac{J_{\rm H}^iJ_{\rm H}^j}{\rho^2}-\frac{{\rm{M_{BH}}}}{r_{\rm H}}\right)\,.
\end{aligned}
\end{equation}
Here $\delta_{ij}$ is the 3D Kronecker delta symbol in spherical coordinates, and we have rewritten $u_\mu^{\rm H}$ using the relation $J^\mu_{\rm H}=\rho u^\mu_{\rm H}$, where $J^{\mu}_{\rm H}$ is the harmonic coordinate counterpart of Eqs.~\eqref{eq:mass_current}. We further note that the DM particles are non-interacting among themselves, hence the internal energy $\Pi$ has been set to $0$.

\begin{figure}[t]
\label{fig:PN_metric_functions}
  \centering
  \includegraphics[scale=0.52]{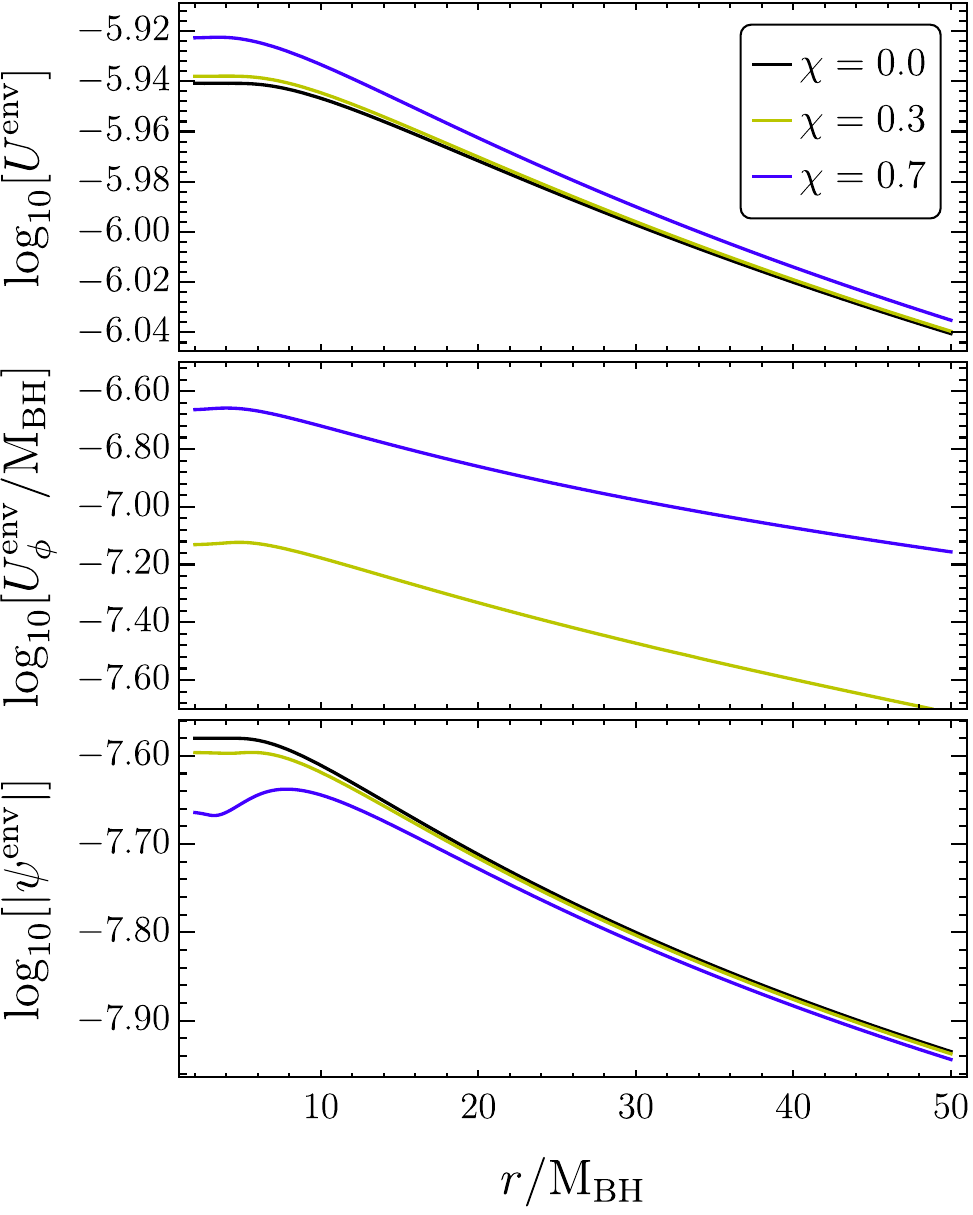}
  \caption{Environmental pieces of the PN metric potentials in Boyer-Lindquist coordinates, computed using Eqs.~\eqref{eq:PN_Poisson_eqs_main_text1} on the equatorial plane. All the environmentally induced PN potentials are at least 5 or 6 orders of magnitude smaller than the corresponding Kerr potentials, regardless of the choice of BH spin.}
\end{figure}

We solve the Eqs.~\eqref{eq:PN_Poisson_eqs_main_text1} numerically via a direct integration scheme, and then transform back to the Boyer-Lindquist coordinates from the harmonic coordinates (see Appendix \ref{app:PN_expansion} for more details) to obtain the metric changes due to the environment. Results of this procedure are shown in Fig.~\ref{fig:PN_metric_functions}, for a DM spike computed with ``Milky-Way like'' parameters, and for several values of the dimensionless spin parameter of the central BH. 

As evident from Fig.~\ref{fig:PN_metric_functions}, the effect of the environment is most pronounced for the $g_{tt}$ component of the metric, in particular $U^{\rm env}$, and for higher values of $\chi$. At smaller values of $(r/{\rm{M_{BH}}})$, the metric contributions due to the environment are the largest for the quantity $U$ appearing in both $g_{tt}$ and $g_{ij}$ components, but at most reach $\sim 10^{-6}$. The contribution of the environment to the other metric coefficients in the PN expansion are even smaller. As such, it is safe to assume that the influence of the Kerr BH on the background is much greater than the environment's influence, and hence we may consider the environment to be a perturbative correction.

Now, we turn to the problem of how the above change in the metric due to the environment affects the orbital dephasing. We first write down the Lagrangian for particles moving through the Kerr+environment spacetime as $\mathcal{L}=(1/2)g_{\mu \nu}u^\mu u^\nu$, where $u^\mu=(\dd x^\mu/\dd \tau)$ is the four-velocity, and $\tau$ is the particle's proper time. Since our background for the Kerr+environment system is both stationary and axisymmetric, the Lagrangian admits two constants of motion, the energy $E$ and angular momentum about the $z$-axis $L_z$:
\begin{equation}
\label{eq:orbital_energy_and_angular_momentum_main_text}
E=-\frac{\pd \mathcal{L}}{\pd u^t}\,,
\qquad
L_z=\frac{\pd \mathcal{L}}{\pd u^\phi}\,.
\end{equation}
For particles orbiting on the equatorial plane, $\theta=\pi/2$ and hence $u^\theta=0$. These conditions alongside Eq.~\eqref{eq:orbital_energy_and_angular_momentum_main_text} allow us to express the Lagrangian in terms of $r$, the radial component of four-velocity $u^r$, and the constants of motion $(E,L_z)$. Massive particles obey the relation $2\mathcal{L}=-1$, and circular orbits require us to set $(\dd r/\dd \tau)=(\dd^2 r/\dd \tau^2)=0$. Using these conditions, we obtain the orbital energy $E_{\rm orbit}$ and orbital angular momentum $L_{z, \rm orbit}$ associated with circular, equatorial orbits as functions of the radius, whose exact expressions are given in Appendix \ref{app:modified_geodesics}. 

With the expressions for $E_{\rm orbit}$ and $L_{z, \rm orbit}$, we may write $u^{t}$ and $u^{\phi}$ as functions of the orbital radius $r$,
and compute the orbital frequency as $\Omega_{\rm orbit}=(\dd\phi/\dd t)=u^\phi/u^t$. To linear order in $\epsilon$, we obtain 
\begin{equation}
\label{eq:modified_orbital_frequency_main_text}
\Omega_{\rm orbit}=\Omega_0-\frac{\epsilon\sqrt{r}}{4{\rm{M_{\rm BH}}}^{3/2}} \left( g_{tt}'^{\text{env}}+2\Omega_0 g_{t \phi}'^{\text{env}}+\Omega_0^2 g_{\phi \phi}'^{\text{env}}\right)\,, 
\end{equation}
where the prime denotes derivative with respect to the Boyer-Lindquist radial coordinate, $\Omega_0=\sqrt{{\rm{M_{\rm BH}}}}/(r^{3/2}+\chi {\rm{M_{\rm BH}}}^{3/2})$ is the angular velocity of a particle on a circular orbit in the Kerr geometry, and the metric components $g_{\mu\nu}^{\text{env}}$ are evaluated on the equatorial plane. 

Lastly, to assess the effect of the environmental perturbation to the metric on binary inspirals, we use a modified form of Eq.~\eqref{numbercycle} to compute the number of orbital cycles built up as a secondary moves from some initial radius, $r$, to $r_{\rm ISCO}$:
\begin{equation}
\label{eq:Ncycles_function_of_r}
\mathcal{N}_{\rm cycle}(r)=
\frac{1}{\pi}\int_{r}^{r_{\text{ISCO}}} \frac{\Omega_{\rm orbit}(r')}{\dot{E}_{\text{DF}}+\dot{E}_{\text{GW}}}\frac{\dd E_{\rm orbit}(r')}{\dd r'}\dd r'.
\end{equation}

To compute the number of cycles using Eq.~\eqref{eq:Ncycles_function_of_r}, we use the vacuum GW fluxes obtained using the \texttt{Teukolsky} package in the \texttt{Black Hole Perturbation Toolkit}~\cite{BHPToolkit,wardell_2025_14788956}, with $\dot{E}_{\rm DF}$ given by Eq.~\eqref{force_DF}, and the modified orbital frequency Eq.~\eqref{eq:modified_orbital_frequency_main_text}. We also take care of the fact that the ISCO radius is modified by the presence of matter.

Our treatment of the orbital properties, e.g. the orbital frequency, energy and $z$ component of angular momentum, is fully relativistic up to $\mathcal{O}(\epsilon)$, since in these computations we keep $g_{\mu\nu}^{\rm env}$ generic. To obtain fully relativistic dephasing results up to $\mathcal{O}(\epsilon)$, one should ideally expand the Einstein equations in powers of $\epsilon$ and solve them for the perturbed metric functions. Since our main goal is to estimate the impact of the changing background metric on the dephasing, we instead take the 1PN metric in Eq.~\eqref{eq:PN_metric_main_text} as a proxy for $g_{\mu \nu}^{\rm env}$ in our computation.

\begin{figure}[t]
\label{fig:PN_dephasing}
  \centering
  \includegraphics[scale=0.62]{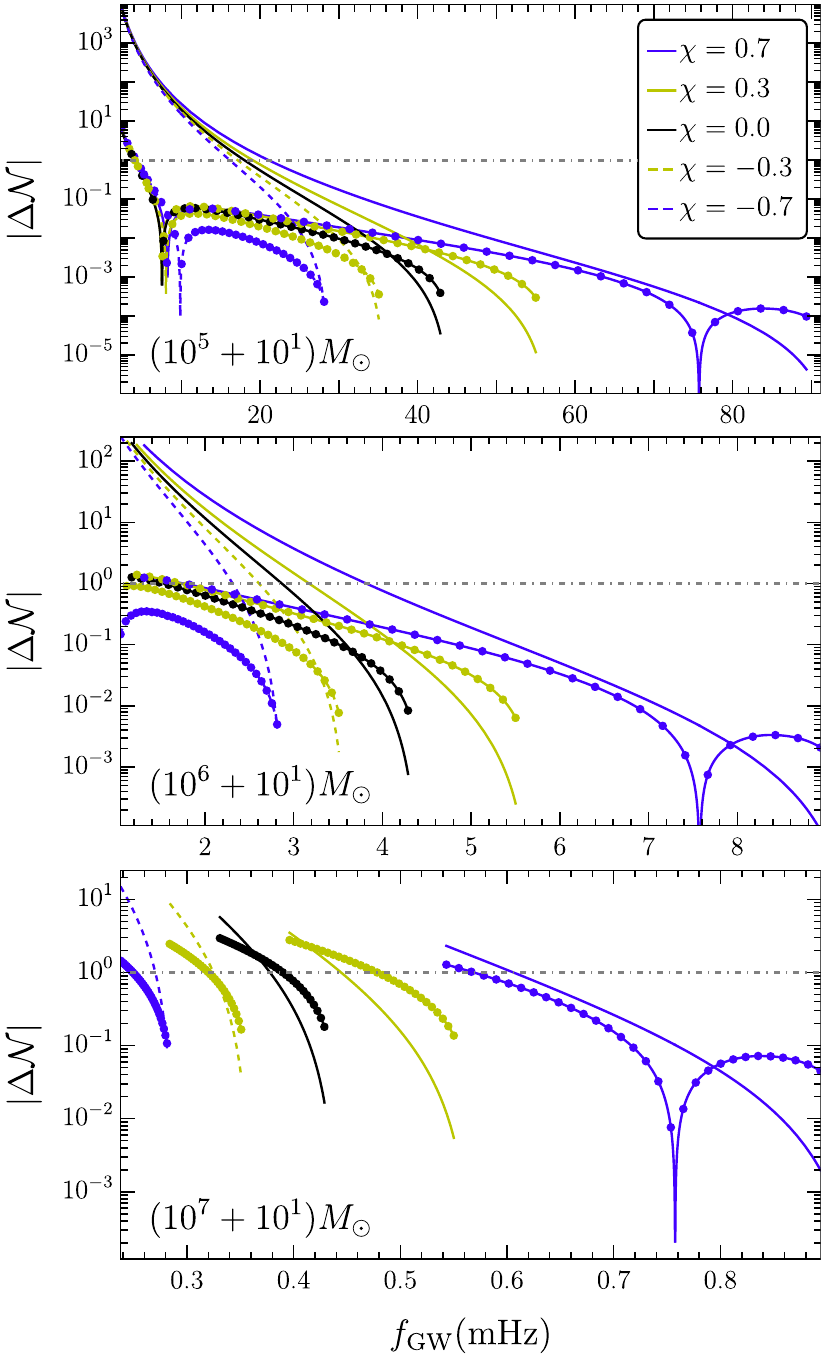}
  \caption{Dephasing contributions for various systems, as a function of the starting GW frequency. Lines with markers refer to the contribution from DF, and lines without markers to the contribution from the environmentally modified metric. Each curve assumes we start at an initial frequency and evolve to the ISCO. For each system, we take the minimum starting frequency to be $\max[f_{10\text{yr}}$,$0.1$mHz], where $f_{10\text{yr}}$ is the starting GW frequency corresponding to a $10$yr inspiral, and $0.1$mHz corresponds to the lower end of the LISA sensitivity. We indicate the $\Delta \mathcal{N}=1$ threshold with a gray dot-dashed line.}
\end{figure}

To isolate the contributions of the metric modifications on dephasing, we first compute the number of cycles including both the DF and metric modifications, 
denoted by $\mathcal{N}_{\rm DF+met}$. We then compute the number of cycles including only the DF (taking $\epsilon\rightarrow0$), $\mathcal{N}_{\rm DF}$, and isolate the dephasing due to the metric change as  $\Delta\mathcal{N}_{\rm met}=|\mathcal{N}_{\rm DF+met}-\mathcal{N}_{\rm DF}|$. Similarly, to isolate the contribution to the dephasing from DF, we take $\Delta\mathcal{N}_{\rm DF}=|\mathcal{N}_{\rm DF+met}-\mathcal{N}_{\rm met}|$. We compare these quantities in Fig.~\ref{fig:PN_dephasing} for a variety of BH masses and spins. As representative values, we set the halo mass and scale radius to be ${\rm{M_{halo}}}=10^{12}M_\odot$ and ${\rm{r_s}}=20\text{kpc}$, and scale the background density $\rho$ with the primary BH mass using Eq.~\eqref{eq:spike_fit}. While we restrict to these halo parameters, we expect similar results for other physically realistic halo configurations. We find that, for systems with ${\rm{M_{BH}}}\in(10^5M_\odot-10^6M_\odot)$, the DF effect is far more important than the metric contributions to the dephasing. Further, $\Delta\mathcal{N}_{\rm met}$ remains below $1$ unless we observe these systems for $\gtrsim8$yrs. This is, however, not the case for ${\rm{M_{BH}}}\sim10^7M_\odot$ systems. While the metric's contribution is often below $\Delta\mathcal{N}_{\rm met}=1$, the DF effect is comparable to the effect of the metric change to the $1\%-10\%$ level. This validates the probe limit approximation taken in our study for primary BHs in the mass range ${\rm M_{BH}} < 10^7 M_\odot$, as discussed in the previous section. We only show $\mu=10M_\odot$ scenarios in Fig.~\ref{fig:PN_dephasing}, but have checked that we get qualitatively similar results for $\mu=10^2M_\odot$ systems as well.

\begin{figure}[t]
\centering
\includegraphics[width=\columnwidth]{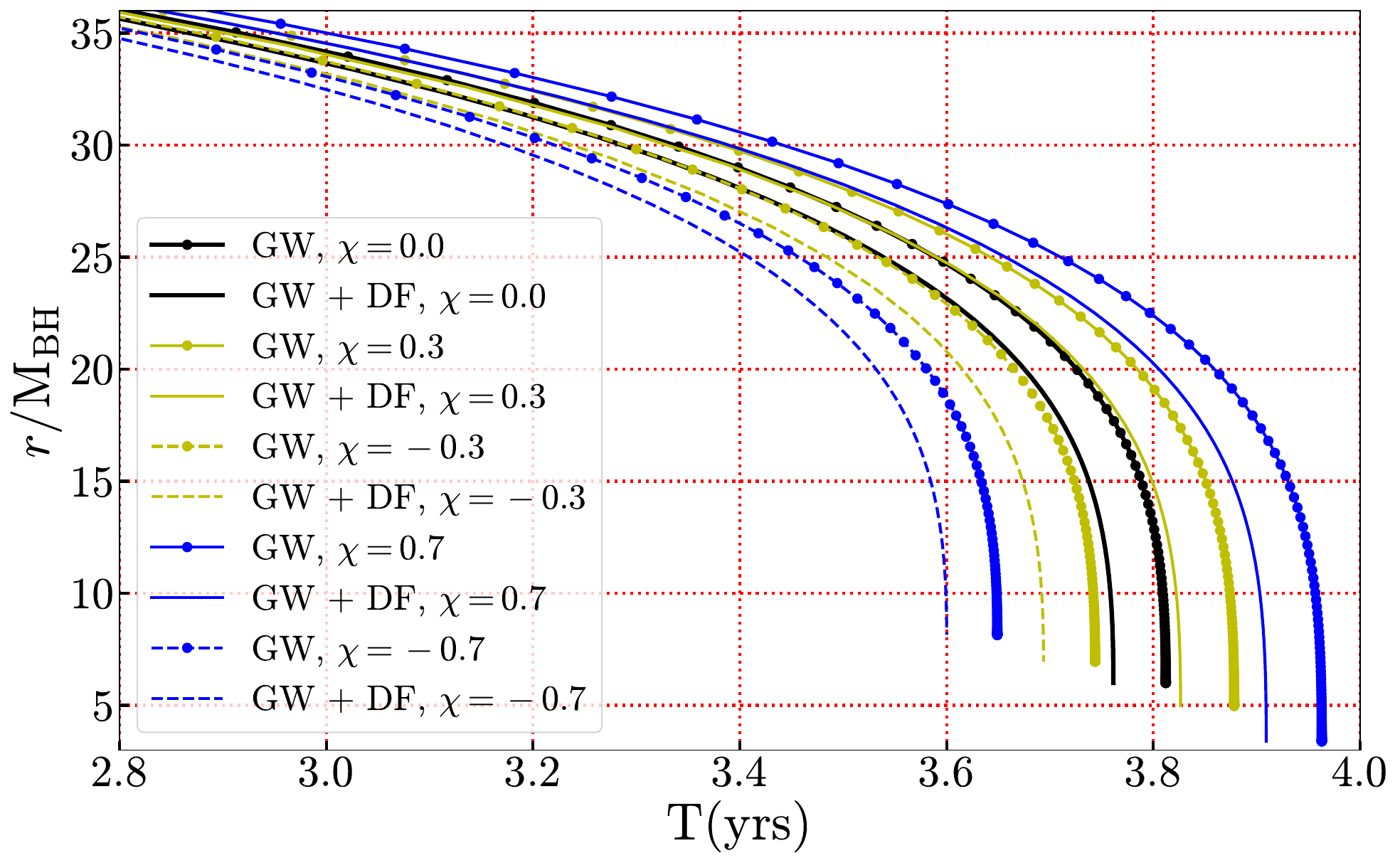}
\caption{Inspiral trajectories of binary systems with and without DF, for representative choices of the dimensionless spin parameter. We take the EMRI to be a $(10^{5}+50)\msun$ system with all orbits starting at an initial radius of $r\ \rm{=50\,M_{BH}}$, with DM halo mass $\rm{M}_{\rm{halo}}=10^{12}$$\msun$ and scale radius $\rm{r_s}=30$kpc. Each curve terminates when the secondary reaches the ISCO radius.}
\label{fig:trajectories}
\end{figure}

\section{Results}\label{sec:results}

\begin{figure*}[htp]
    \centering
    \includegraphics[width=\textwidth]{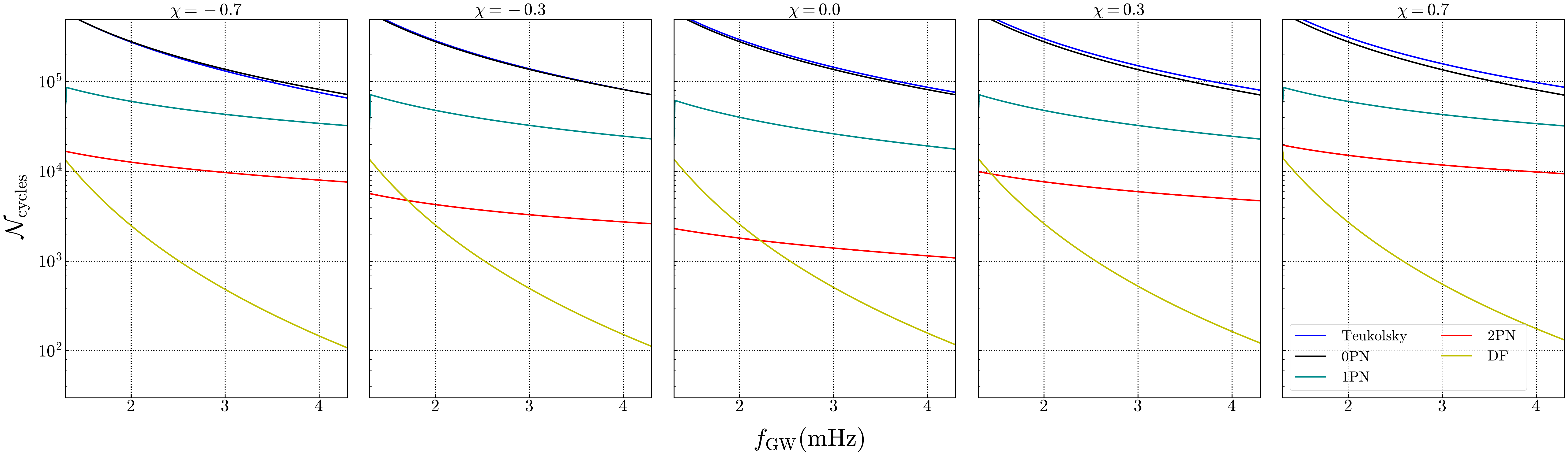}
    \caption{Number of cycles that accumulate as the secondary inspirals from a given starting frequency to the ISCO, for a $(10^5+50)M_\odot$ EMRI system surrounded by a $\rm{M}_{halo} = 10^{12}$$\msun$, $\rm{r_s} = 30 \rm{kpc}$ DM spike. We isolate the contributions to the cycling from different sources of energy loss: the fully relativistic Teukolsky flux (blue), 0PN (black), 1PN (dark cyan), and 2PN (red) contributions to the GW flux, and energy loss due to DF alone (yellow). We have considered five different choices for the dimensionless spin parameter $\rm{\chi}=0.0, \pm 0.3$ and $\pm 0.7$.
    }
    \label{fig:Ncycle_comparison}
\end{figure*}

\begin{figure*}[ht]
    \centering
    \subfigure{\includegraphics[scale=0.30]{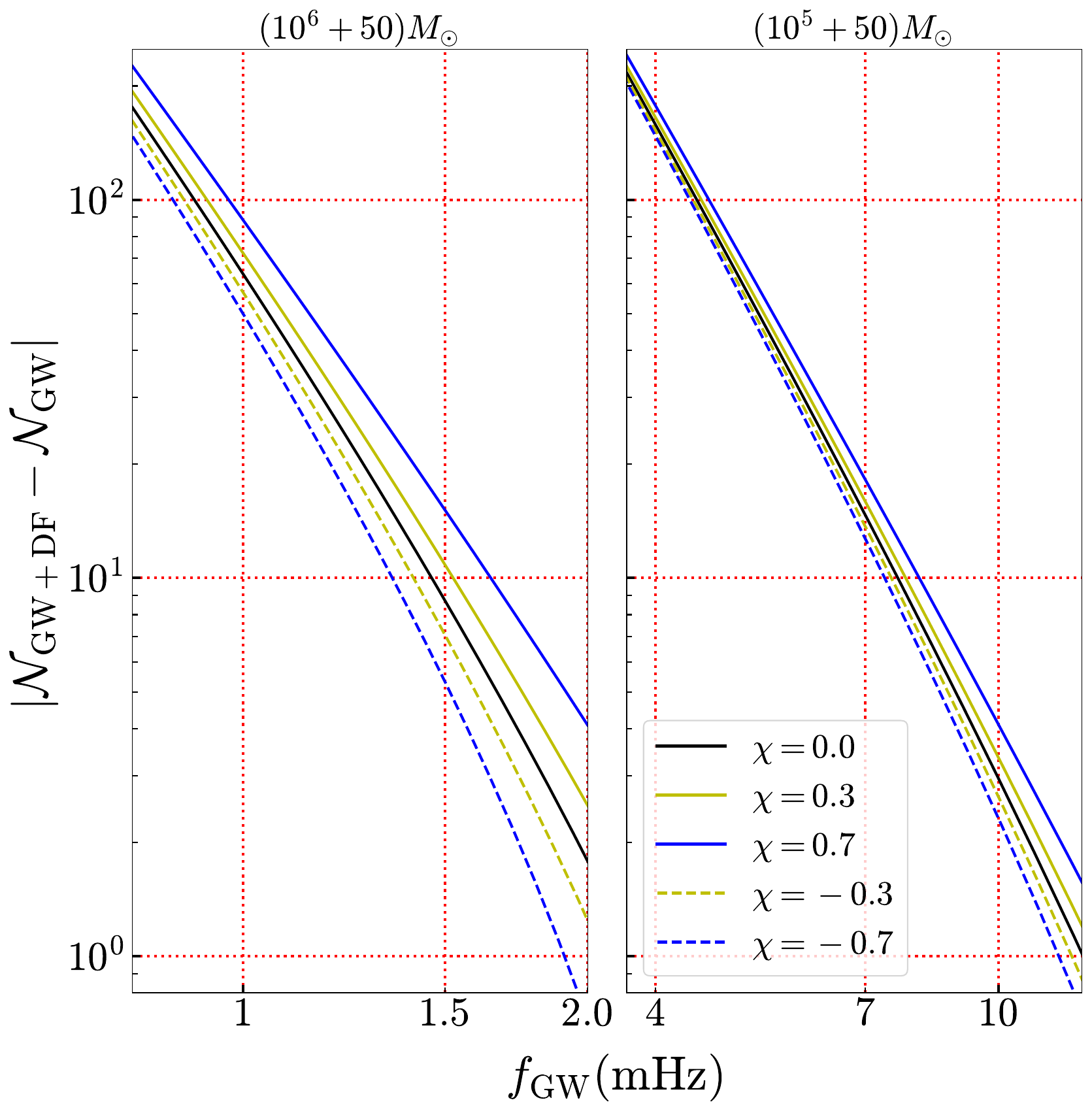}
    \label{fig:dephasing1a}}
    \subfigure{\includegraphics[scale=0.30]{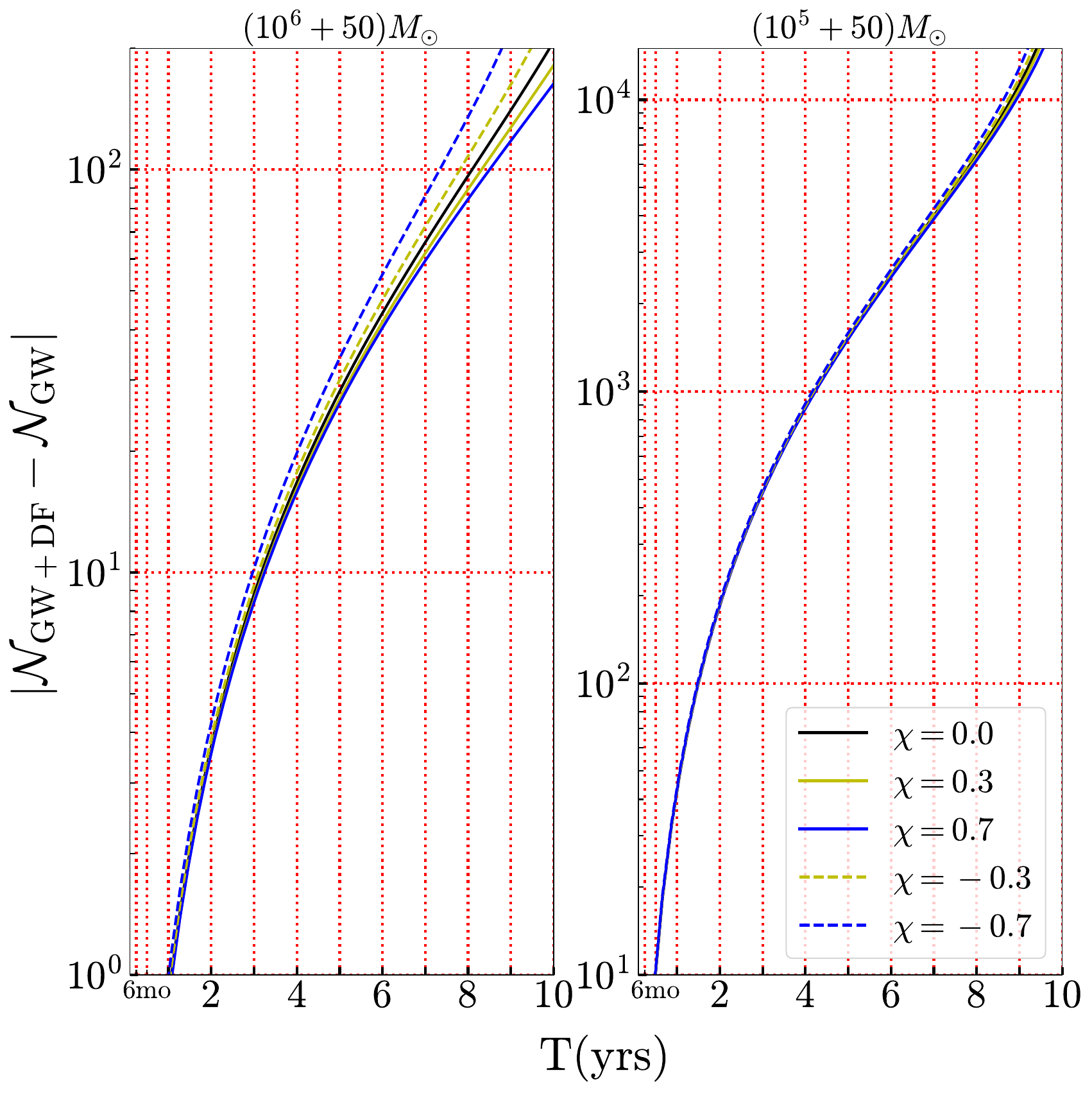} \label{fig:dephasing1b}}
    \caption{Dephasing due to DF as a function of GW frequency at the start of observation (two plots on the left) and as a function of time of inspiral (two plots on the right), for two different mass ratios $(10^6+50)\msun$ and $(10^5+50)\msun$. In both plots,  the solid blue, yellow, and black lines correspond to spin values of $\chi = 0.7$, $0.3$, and $0.0$ respectively, while the dashed yellow and blue lines correspond to $\chi = -0.3$ and $-0.7$. The \dm\ spike
    parameters are set to representative values of  $\rm{M}_{\rm{halo}} =10^{12} $$\msun$ and $\rm{r_{s} = 30}$ kpc.}    
    \label{fig:dephasing1}    
\end{figure*}

In this section, we summarize the main effects of the rotating \dm\ spike profiles on the gravitational waveform. As emphasized before, we start by computing inspiral trajectories for binary systems with \texttt{FEW}, modifying the orbital energy and angular momentum loss as in Eqs.~\eqref{energy balance} and~\eqref{ang_mom balance}. The result of such an analysis for a $(10^{5}+50)M_{\odot}$ EMRI system is presented in Fig.~\ref{fig:trajectories}, where we show how the Boyer-Lindquist radial coordinate $r$ (normalized with respect to the primary BH mass) evolves with time, for various choices of the dimensionless spin parameter $\chi$ and with or without DF. 
As expected, Fig.~\ref{fig:trajectories} shows that for all spin values, the DF adds to the energy loss of the binary, thereby accelerating the inspiral towards the ISCO. Moreover, the inspiral time varies significantly between prograde and retrograde orbits. For example, the retrograde orbit around a spinning BH with dimensionless spin parameter $\chi=0.7$ takes about 3.6 years to inspiral, while a prograde orbit takes almost 3.9 years. This effect is primarily due to the different ISCO radii between the two cases: the retrograde orbits have a much larger ISCO radius compared to the prograde orbits, and hence they reach the ISCO in less time. This will have direct implications for the discussion in this section.

Next, in Fig.~\ref{fig:Ncycle_comparison} we compute various contributions to the number of cycles  from the DF and from different PN orders in the GW energy flux to gauge their relative importance. We also highlight the influence of the central BH's rotation, which leads to both a modified DM spike distribution and to different GW emission at higher PN orders. For this purpose, we have plotted the number of cycles for five different choices of the (dimensionless) spin values, $\chi=0.0, \pm0.3$ and $\pm0.7$. 
As expected, the relativistic Teukolsky flux is dominated by the 0PN contribution (quadrupolar flux). The 0PN flux slightly overestimates the number of cycles in retrograde inspirals, while underestimating the prograde ones. This discrepancy increases for higher spin values and higher starting frequency of observation. This is expected, as the higher PN terms start to contribute more with increasing spin and smaller initial radius (or, equivalently, larger starting frequency) of the inspiral. The DF, on the other hand, being a negative PN effect, contributes very little for larger starting frequency (or, smaller starting radius) and always remains subdominant to both the 1PN and the 2PN contributions. However, for lower starting frequency of the inspiral (e.g., $\simeq 1.5\,\rm{mHz}$), the DF becomes comparable with the 2PN effect for $\chi=0.3$, and even becomes dominant over it for the zero spin and retrograde cases. Another interesting feature of Fig.~\ref{fig:Ncycle_comparison} is that, for both the prograde and retrograde cases, the effect of DF becomes subdominant to the 2PN term as the BH spin magnitude increases, even for the same starting observation frequency. In other words, while a larger spin enhances the DF effect due to the increased DM spike densities, the increased spin enhances the higher order PN flux terms more. In the non-rotating case, in particular, the DF dominates significantly over the 2PN term at small starting frequencies. 

In Fig.~\ref{fig:dephasing1} we consider again the five cases $\chi=0$, $\chi=\pm0.3$, and $\chi=\pm0.7$, and show the results of dephasing between vacuum and non-vacuum inspirals. We isolate the DF's contribution to the dephasing by computing the quantity $|\mathcal{N}_{\rm{GW+DF}}-\mathcal{N}_{\rm{GW}}|$, where $\mathcal{N}_{\rm{GW}}$ is the number of cycles for the inspiral driven only by GW emission, and $\mathcal{N}_{\rm{GW+DF}}$ is the number of cycles for an inspiral driven by both GW emission and DF energy loss. We compute the dephasing in two different ways: first, as a function of initial GW frequency, i.e., the frequency at which the observation is started (leftmost two plots in Fig.~\ref{fig:dephasing1a}); second, as a function of time (rightmost two plots in Fig.~\ref{fig:dephasing1b}). For the plots in Fig.~\ref{fig:dephasing1a} we consider the initial GW frequencies to be inside the LISA sensitivity band, and assume that we observe the system until the secondary reaches the ISCO, whose location can be different depending on particular values of the rotation parameters. The initial frequency is chosen such that the ISCO is reached before a $\rm{10~yr}$ observation period. On the other hand, in Fig.~\ref{fig:dephasing1b} we start the inspirals at a fiducial value of 1.25 mHz (for a $(10^5+50)\msun$ system) and 0.63 mHz (for $(10^6+50)\msun$ system), and evolve the systems forward in time. The fiducial values are chosen such that the SNR of the waveform for the inspiral in the \dm\ environment reaches a value of 20 (i.e., becomes observable) at $\rm{T \sim 4-6~yrs}$, and such that both inspirals take $\rm{T > 10 yrs}$ to reach the ISCO.   

\begin{figure}[t]
    \centering
    \includegraphics[width = \columnwidth]
    {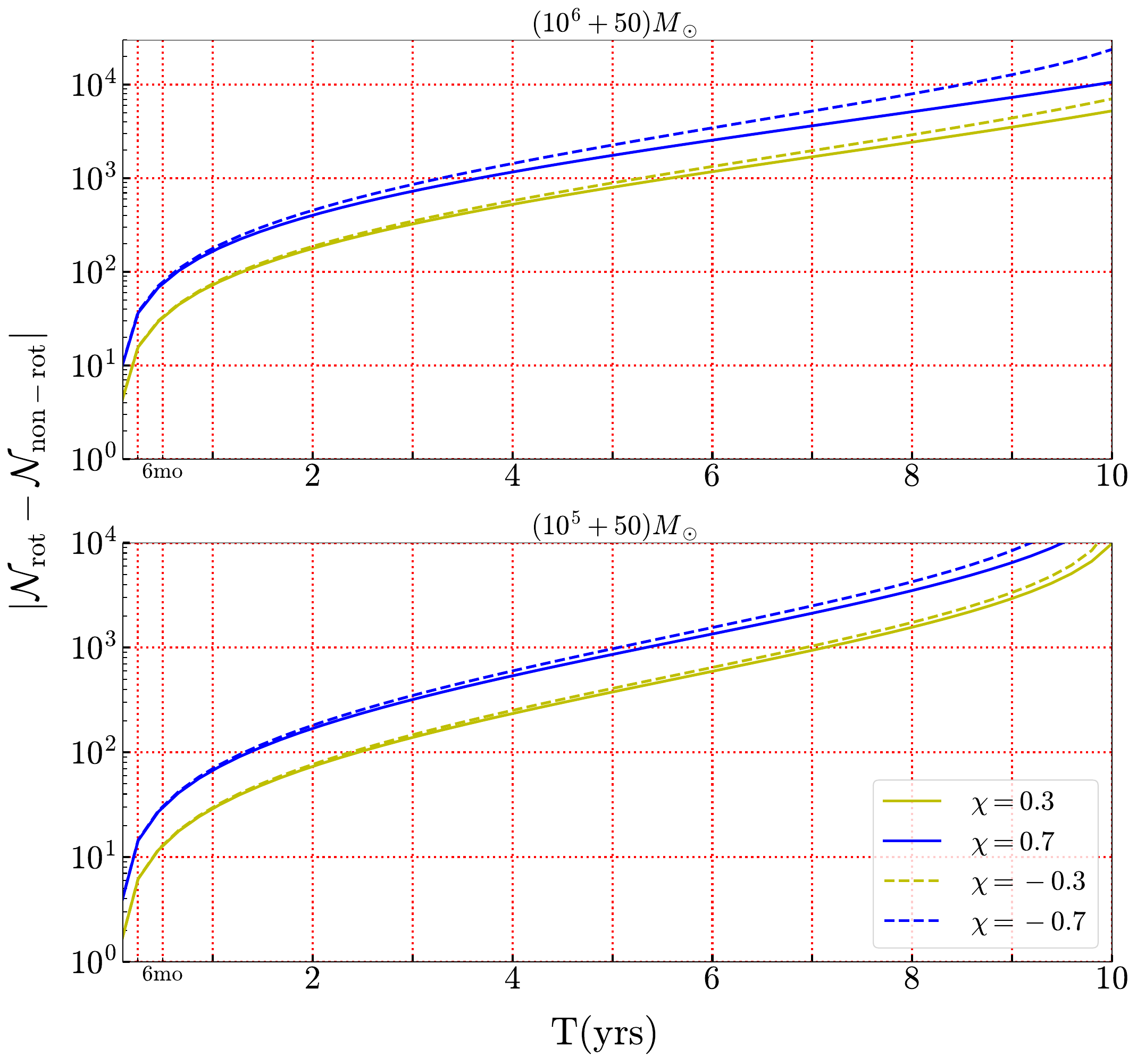}
    \caption{Difference in the number of cycles between rotating and non-rotating EMRIs surrounded by DM spikes, as a function of time of inspiral. We take two different mass ratios and initial observation frequencies --- (a) $f_{\rm ini}=0.63~\text{mHz}$ for the $(10^6+50)\msun$ system, and $f_{\rm ini}=1.25~\text{mHz}$ for the $(10^5+50)\msun$ system. These initial frequencies are chosen such that one obtains an SNR $\sim$ 20 after $T\sim 4$ yrs observation time. The solid blue and yellow lines correspond to the prograde trajectories for $\chi=0.7$ and $0.3$, respectively, while the dashed ones correspond to retrograde trajectories with $\chi=-0.7$ and $-0.3$. The \dm\ spike parameters are the same in both cases with total halo mass ${\rm{M_{halo}}}=10^{12}M_\odot$ and scale factor $\rm{r_s}=30$kpc.}
    \label{fig:dephasing2}
\end{figure}

In Fig.~\ref{fig:dephasing1a}, for both mass ratios, we can see some salient features. The maximum dephasing for a certain starting frequency 
occurs for $\chi = 0.7$, i.e. the prograde case with the highest spin value. The dephasing decreases with decreasing spin as one goes to non-rotating case. The trend continues as one crosses over to negative values of spin (i.e., retrograde orbits), and the dephasing continues to decrease as the spin becomes more negative. This result can be connected with our earlier observations in Fig.~\ref{fig:trajectories}. The retrograde orbits finish their inspiral earlier than the prograde ones, and hence the number of cycles in the total duration of the inspiral is smaller for the retrograde orbits than for the prograde ones. Further, for larger positive spins, the position of the ISCO is much closer to the central BH. As such, the secondary object is able to probe a larger region of the DM spike, and can spend more time moving through the highest density regions of the spikes close to the central BH. One can also see that the dephasing goes down if one increases the starting frequency. This is once again expected, as higher starting frequencies translate to smaller starting radii and shorter inspirals, so the change of cycles does not build up to an appreciable level. On the other hand, in Fig.~\ref{fig:dephasing1b}, the inspiral continues and the dephasing experienced by retrograde orbits due to DF is seemingly higher than the prograde orbits. The difference enhances as the spin of the central BH is increased.

\begin{figure}[t]
    \centering
    \includegraphics[width = \columnwidth]
    {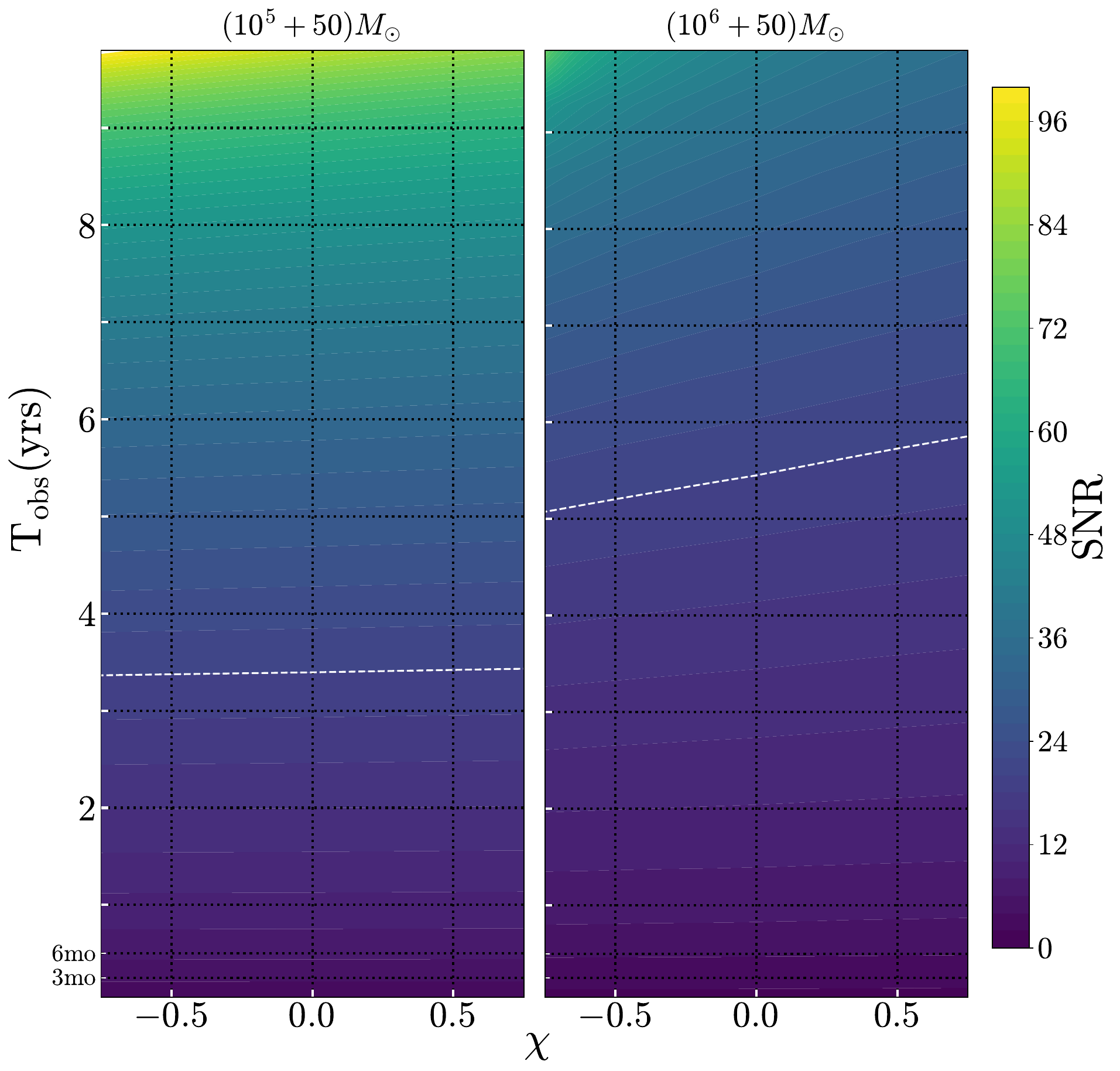}
    \caption{SNR as a function of observation time and central BH spin. For the left plot, we fix the initial frequency at the start of observation to 1.25 mHz for a $(10^{5}+50)\msun$ system. On the right, we fix the initial frequency to 0.63 mHz for a $(10^{6}+50)\msun$ system. The DM spike is parametrized by the total halo mass $ {\rm{M_{halo}}}=10^{12}$$\msun$ and scale factor $\rm{r_s}=30~kpc$. The white dashed line corresponds to an SNR of 20, at which point the waveform is considered to be observable.
    }
    \label{fig:SNR_Tobs}
\end{figure}

\begin{figure*}[t]
\label{fig:Mismatch_vac_comp}
  \centering
  \includegraphics[scale=0.38]{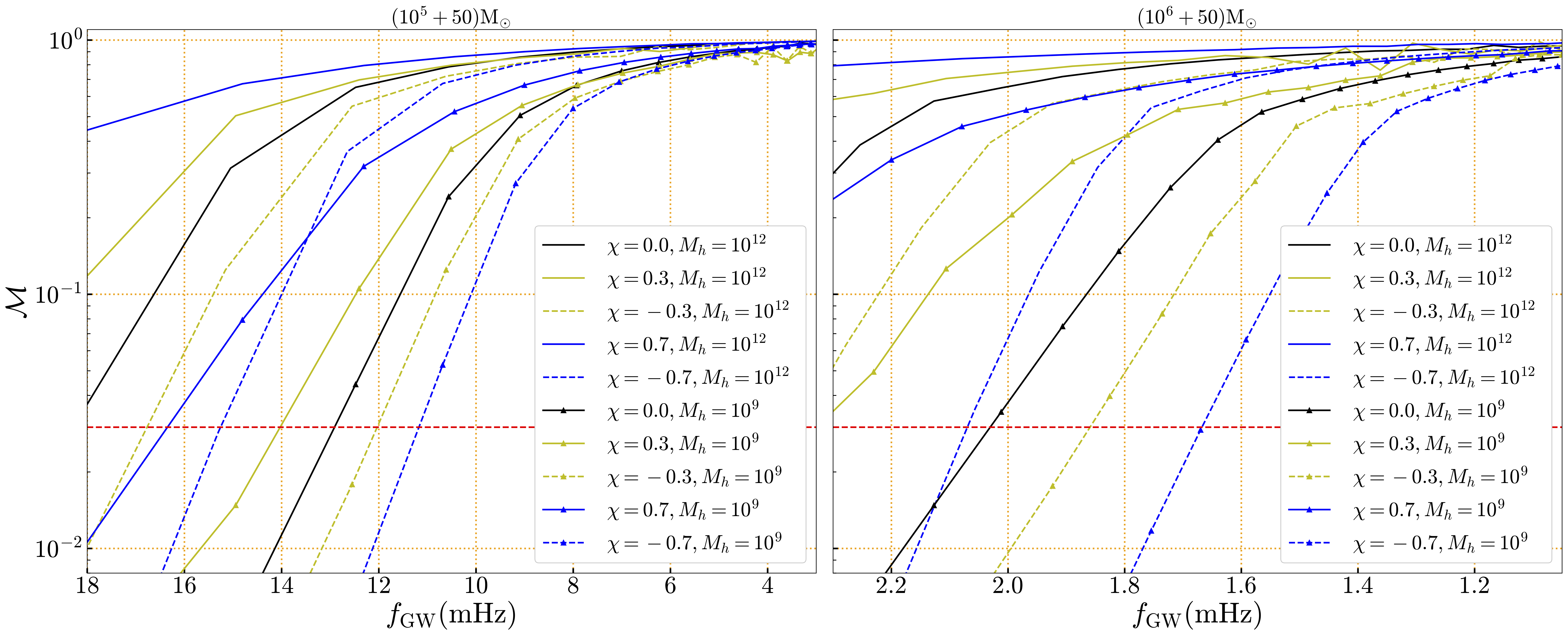}
  \caption{Mismatch between waveforms in vacuum and in a DM spike environment, as system evolves from a given initial GW frequency to the ISCO. We consider two different EMRIs with masses $(10^{5}+50)\msun$ and $(10^{6}+50)\msun$, and change the central BH spin alongside the DM spike mass. The mismatch for prograde(retrograde) orbits are denoted by solid(dashed) lines. For the spike, we consider two different halo masses, $M_h = {\rm{M_{halo}}}=10^{12}\msun$ and $10^{9}\msun$ (denoted by the triangle marked curves), with the same scale radius $\rm{r_s}=30~kpc$. }
\end{figure*}

Having considered the effect of DF between vacuum and non-vacuum inspirals, we next examine how the non-vacuum GW phase is modified due to rotation. In Fig.~\ref{fig:dephasing2}, we plot the absolute value of the difference between the number of cycles in the rotating case ($\mathcal{N}_{\rm rot}$) to that of the non-rotating case ($\mathcal{N}_{\rm non-rot}$) as a function of time. We again consider $(10^{6} +50)\msun$ and $(10^{5}+50)\msun$ systems, and again set the fiducial starting GW frequencies to $f_{\text{ini}}=0.63~\text{mHz}$ and $f_{\rm ini}=1.25~\text{mHz}$, respectively. For both the rotating and non-rotating cases, the inspirals are driven by both GW emission and DF. Given the above setup, Fig.~\ref{fig:dephasing2} primarily captures the dephasing effect arising from the difference in the \dm\ spike profiles for the various spins. We see that, irrespective of prograde or retrograde trajectories, the dephasing in the $\chi=\pm 0.7$ systems dominates over the dephasing from $\chi=\pm 0.3$ systems. This may be explained by noting that the \dm\ spike densities increase as the magnitude of $\chi$ is increased, with the spike taking its lowest density at $\chi=0$. We further note that, in all the cases, the retrograde orbits experience a larger number of cycles than the prograde ones.  

\begin{figure}[ht]
    \centering
    \includegraphics[width = \columnwidth]
    {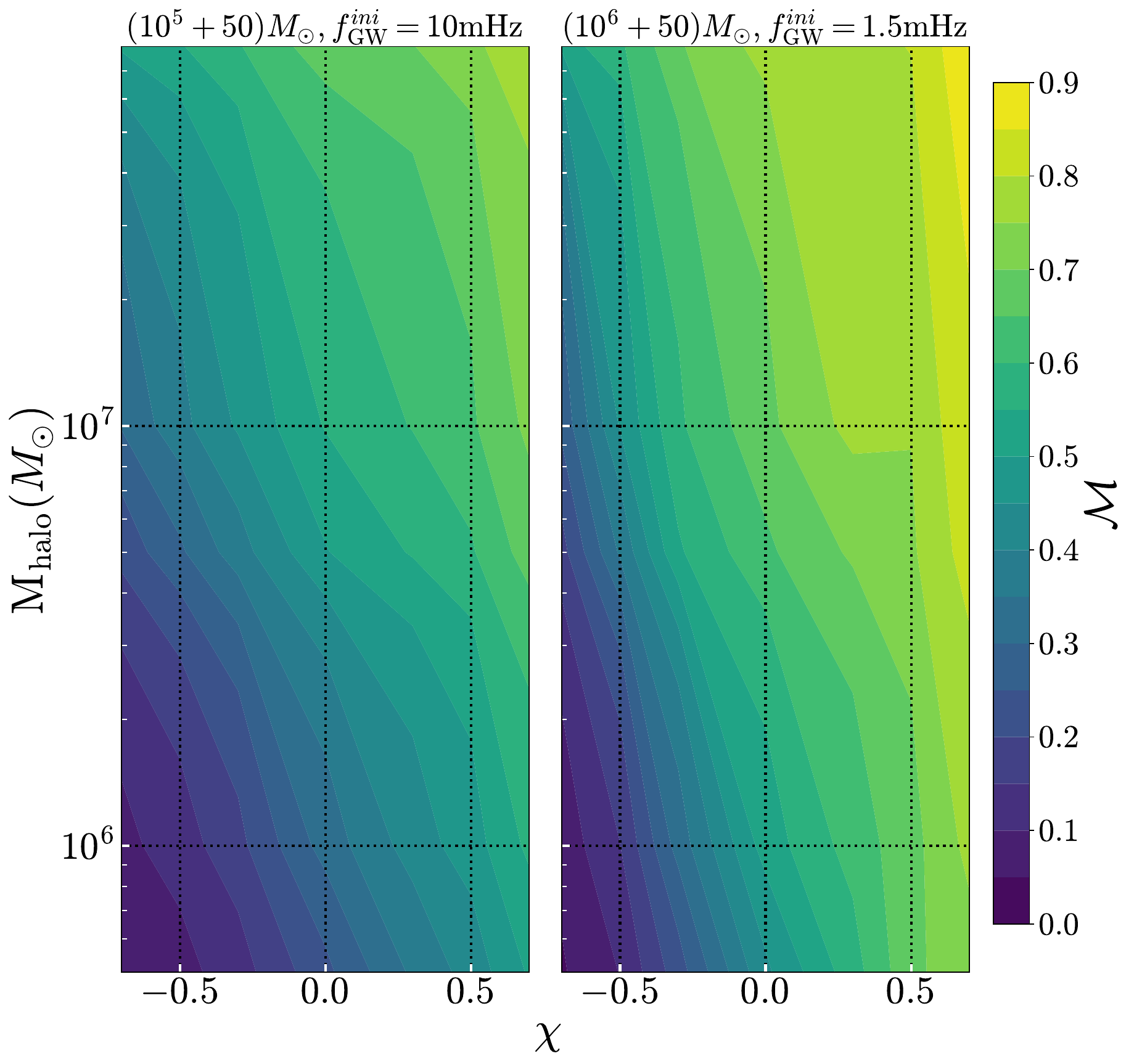}
    \caption{Mismatch between gravitational waveforms in vacuum and in the DM environment, as a function of the rotation parameter of the central BH and the mass of the DM halo. We consider $(10^{5}+50)\msun$ and $(10^{6}+50)\msun$ EMRI configurations with starting frequencies of $10$ mHz and $1.5$ mHz, respectively. The scale factor is taken to be $\rm{r_s} = 10 kpc$. We have considered the evolution of inspiral until it reaches ISCO. }
    \label{fig:Mismatch_contour_a_mgal}
\end{figure}

So far, we have only discussed the dephasing of gravitational waveforms due to DM spikes and rotation of the central BH. To better assess the  strength of the signal with DM-induced effects at the detector, in Fig.~\ref{fig:SNR_Tobs} we consider the behavior of the SNR, defined in Eq.~\eqref{SNR_defn}. We show the buildup of SNR for non-vacuum waveforms (i.e., inspiral trajectories driven by both DF and GW emission) for different values of the spin parameter and the time of observation considering the same fiducial starting GW frequencies as before, $f_{\text{ini}}=0.63~\text{mHz}$ and $f_{\rm ini}=1.25~\text{mHz}$ for $(10^{6} +50)\msun$ and $(10^{5}+50)\msun$ systems, respectively. As expected, the SNR increases with increasing observation time, with an SNR value of $20$ achieved in $\sim 3.5~\rm{yrs}$ for the $(10^{5}+50)\msun$ system. For the $(10^{5}+50)\msun$ system, the effect of the spin parameter is small. This contrasts with the $(10^{6}+50)\msun$ system, which shows a much higher sensitivity to spin. For example, the retrograde orbits reach the $\rm{SNR \sim 20}$ threshold almost a full year earlier than their prograde counterpart, with an average of $5.5$ yrs of observation time. Figure~\ref{fig:SNR_Tobs} shows that retrograde orbits experiencing DM-induced DF may become louder than their prograde counterparts if we are able to continuously observe GWs from environmentally impacted EMRIs.

Next, to make more concrete statements about the detectability of DM induced effects, we perform mismatch analyses between vacuum and non-vacuum gravitational waveforms. We show these mismatches in Fig.~\ref{fig:Mismatch_vac_comp} as a function of initial frequency at the start of observation, for $10^{12}$$\msun$ and $10^{9}$$\msun$ DM halo masses, and assuming the secondary evolves to the ISCO. As expected, with decreasing initial frequency (corresponding to a larger starting radius/longer observation time) the mismatch increases, and ultimately becomes greater than the critical value of $\mathcal{M} \sim 0.03$ above which the two waveforms are usually considered to be distinguishable~\cite{Damour:1997ub,Lindblom:2008cm,Chandramouli:2024vhw}, marked by a dashed red line in Fig.~\ref{fig:Mismatch_vac_comp}. For the $(10^5 +50) \msun$ system, the mismatch is larger than the critical value for all the initial frequencies in the LISA band $(\leq 10\rm{mHz})$ and for both values of the halo mass. For the $(10^{6}+50)\msun$ system, on the other hand, the mismatch becomes appreciable only for very low starting frequency: $\lesssim 2.1$ mHz for halo mass $10^{12}M_{\odot}$, and $\lesssim 1.7$ mHz for halo mass $10^{9}M_{\odot}$. Thus for the $(10^{6}+50)\msun$ system, the mismatch between the waveforms in vacuum and in the DM environment is not always significant enough to be observable. For a given starting frequency, the mismatch is also consistently higher for prograde orbits, as they inspiral for a longer duration compared to the retrograde cases. At higher values of the mismatch, one can observe an oscillating pattern which results from an accidental coherence/decoherence between vacuum and \dm\ environment waveform at high dephasing. These oscillations do not carry any significant physical meaning, as noted and explained in~\cite{Speeney:2022ryg}.

\begin{figure}
    \centering
    \includegraphics[width=\columnwidth]{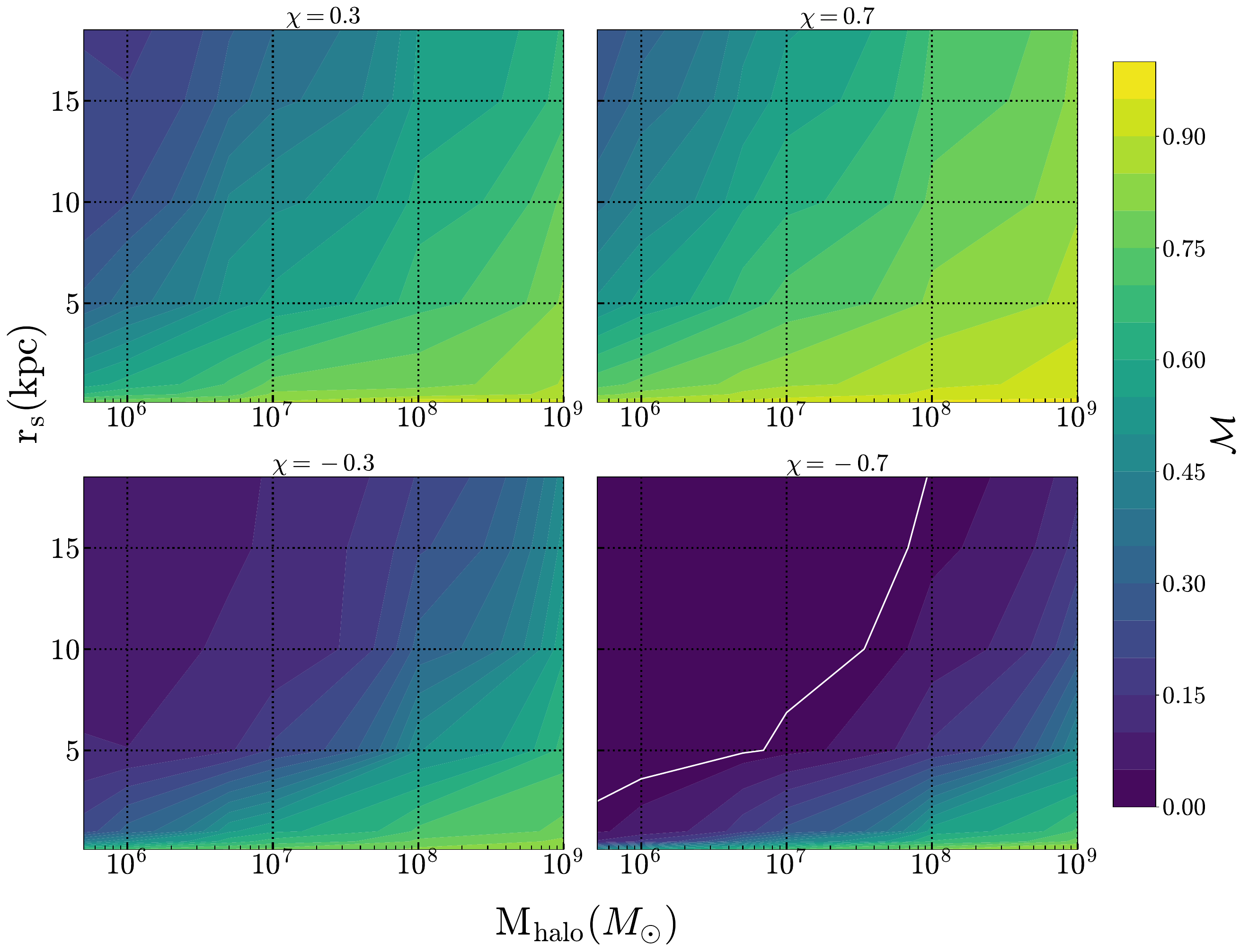}
    \caption{Mismatch as a function of scale radius $\rm{r_s}$ and halo mass ${\rm{M_{halo}}}$ for a $(10^6 + 50) M_\odot$ system. We start the inspirals at  $1.5$ mHz,and continue to evolve until ISCO. We have further takes spins of $\pm0.3$ and $\pm0.7$ as representative values. The white line corresponds to a mismatch value of $0.03$, above which the two waveforms are considered to be distinguishable.}
    \label{fig:Mismatch_contour_rs_mgal}
\end{figure}

Lastly, we extend our previous mismatch computations to study how the halo parameters affect the observability of DM induced effects. With a fixed scale radius $r_{\rm s}=10 {\rm kpc}$, in Fig.~\ref{fig:Mismatch_contour_a_mgal} we plot the mismatch between vacuum and non-vacuum waveforms for various halo masses and BH spins. For the $(10^{5}+50)M_{\odot}$ system, we start the inspiral with an initial frequency of $10$ mHz, while the $(10^{6}+50)M_{\odot}$ system starts at $1.5$ mHz. In all of these cases the mismatch for retrograde orbits is smaller than for prograde orbits, for a given halo mass. Further, for larger halo masses the mismatch is always larger than the threshold value of $0.03$. We further investigate how changing the halo's total mass and scale radius affects the mismatch in Fig.~\ref{fig:Mismatch_contour_rs_mgal}. We show how the mismatch changes for various halo masses and scale radii, choosing $\chi=\pm0.3$ and $\chi=\pm0.7$ as representative spin values. For a given halo mass, the mismatch decreases as we increase the halo's scale radius. This is easily explained by examining the halo parameter scalings in Eq.~\eqref{eq:spike_fit}, and noting that an increased scale radius leads to a more diffuse DM halo. Further, for prograde orbits the mismatch increases with $|\chi|$, while for retrograde orbits, the mismatch decreases with $|\chi|$. Once again, this feature can be attributed to the different prograde and retrograde values of the ISCO: the prograde orbits have longer inspirals, and hence have more cycles to build up differences between the vacuum and non-vacuum waveforms.

\section{Concluding remarks} 

We have investigated the effect of DM spikes on EMRI systems and how the addition of spin affects GW observables by utilizing a fully relativistic framework for the spike~\cite{Ferrer:2017xwm} and for the binary evolution~\cite{Katz:2021yft,michael_l_katz_2023_8190418}. The DM spikes become more pronounced as the spin of the central supermassive BH increases, and hence dephasing effects due to DM spikes are also larger. Considering inspirals which evolve all the way to the ISCO, we showed that prograde inspirals are more impacted by DM effects, primarily due to their smaller ISCO radius compared to retrograde orbits. For inspirals evolving for fixed time durations, however, the retrograde orbits experience a larger dephasing, which increases with increasing BH spin. Through dephasing and mismatch computations, we further showed that the central BH's spin can enhance the detectability of DM effects with LISA, and showed how this detectability can vary with the DM halo parameters. For example, for a $(10^{6}+50)\msun$ EMRI system, it takes $\sim 5\rm{yrs}$ for a gravitational waveform from a retrograde EMRI in a \dm\ environment with the spin of the central BH being $0.8$ to become detectable (i.e., $\rm{SNR} =20$), while for prograde EMRI, the same SNR is reached in $\sim 6 \rm{yrs}$. These results need to be corroborated with a parameter estimation analysis, that we will present in future work.

Our results imply that rotational effects must be carefully accounted for when studying EMRIs with additional environmental effects. In particular, rotation affects the DM distribution in the environment, not just the binary dynamics. Incorporating both effects simultaneously is essential, as any parameter estimation study will be biased due to the significant GW dephasing caused by the environment. Further, mismodeling such effects (for example, including spin only in the dynamics) can result in significant biases in GW parameter estimation.

In summary, the rotation of the central BH significantly affects the detectability of the DM-induced effects in gravitational waveforms. Via SNR and mismatch computations, we have investigated how the observability of DM spikes through the DF effect can change as we vary the BH spin, the DM halo mass, and the DM halo scale radius. For a fixed scale radius, we found that rotation allows for lower mass DM spikes to be distinguishable from vacuum, with the best observability prospects happening for high-spin, prograde orbits. When both mass and radius are allowed to vary, we found that high-spin, prograde orbits are again optimal for probing more diffuse halos, with retrograde orbits requiring a more compact, more massive halo to have observable effects.

For completeness, we have also shown that realistically diffuse DM halos have a negligible effect on the background metric for systems with a central BH mass ${\rm{M_{BH}}}\lesssim10^7M_\odot$. We considered the DM environment to act as a perturbative correction to the Kerr metric, and found the modified background in a PN framework. By solving the relevant Poisson equations in this context, we have shown that the correction to the metric due to the DM environment is roughly 5 orders of magnitude smaller than the background Kerr contribution. Using this modified metric, we have computed orbital dephasings for various primary BH masses and spins, and isolated dephasing contributions coming from the DF and from changes in the metric. For $\sim (10^5-10^6)M_\odot$ primary mass systems, the dephasing induced by the modified metric is at least two orders of magnitude smaller than the DF-induced dephasing, showing that the DF effect is typically much more important. Further, the dephasing due to the modified metric remains below the $|\Delta\mathcal{N}|=1$ threshold, implying that our prior dephasing and mismatch results are not significantly impacted by ignoring changes in the metric. However, for central BH masses ($\gtrsim10^{7}\,M_{\odot}$), the dephasing due to the modified metric becomes comparable to that of the DF, and hence it is no longer safe to ignore the DM's impact on the metric for these systems. 

Our work lends itself to several follow-up studies. First, a proper parameter estimation study including the effects of DF and the relativistic spike distributions around a rotating BH would offer a more complete picture of the observability and detectability of such systems. Some studies have attempted to use GW signals to estimate environmental parameters in spherically symmetric systems~\cite{Cole:2022yzw,Cole:2022ucw}, yet it remains an open question whether this will be possible for generically rotating EMRIs, and to what degree parameter degeneracies may be exacerbated by the environment. Secondly, our work is limited to primary BHs in the mass range of $(10^{5}- 10^{6}) \msun$. The lower bound stems from our assumption of a stationary DM spike~\cite{Kavanagh:2020cfn}, with the upper bound coming from our neglect of the back-reaction of the DM spikes on the metric. For precision GW measurements with LISA and other next-generation detectors, and to incorporate EMRIs of a broader mass ratio, we need to self-consistently model the back-reacted Kerr metric due to the rotating DM spikes. We are currently working in both of these directions. Besides this, one might relax the condition that the secondary BH's velocity and the DF force are in opposite directions. This gives rise to the gravitational Magnus effect for rotating EMRIs, which can induce out-of-plane motion and give rise to very rich physics. The effect of eccentricity is also important, in light of conflicting claims regarding the increase or decrease of eccentricity in the presence of environmental effects~\cite{Becker:2021ivq,Yue:2019ozq}. Incorporating eccentric orbits in our rotating DM spike configurations may be an interesting extension of our work, and may offer further insights into claims of orbital circularization.

\acknowledgments

This work makes use of the Black Hole Perturbation Toolkit~\cite{BHPToolkit}.
The authors would like to thank Augusto Medeiros da Rosa for providing code utilized in this work, and Hassan Khalvati for useful discussions.
E.B. and N.S. are supported by NSF Grants No. AST-2307146, PHY-2207502, PHY-090003 and PHY-20043, by NASA Grant No. 21-ATP21-0010, by the John Templeton Foundation Grant 62840, by the Simons Foundation, and by the Italian Ministry of Foreign Affairs and International Cooperation grant No.~PGR01167.
S.M. thanks the University of South Dakota's High Performance Computing cluster \texttt{Lawrence}, that was used for a significant part of the calculations. Additional numerical work was carried out at the Advanced Research Computing at Hopkins (ARCH) core facility (\url{https://www.arch.jhu.edu/}), which is supported by the National Science Foundation (NSF) grant number No.~OAC-1920103. The research of S.C. is supported by MATRICS (MTR/2023/000049) and Core Research (CRG/2023/000934) Grants from SERB, ANRF, Government of India. S.C. also thanks the local hospitality at ICTS and IUCAA through the associateship program, where a part of this work was done.

\appendix
\section{Post-Newtonian computation of spacetime metric with the DM environment}
\label{app:PN_expansion}

Here, we primarily follow the methodology for building neutron stars in a PN expansion laid out in~\cite{Andersson:2022cax}, adapted to our matter distribution. The goal is to obtain a PN metric which accounts for the central Kerr BH \textit{and} the surrounding DM. In this section, we utilize $\rm{M_{BH}}=1$ units for simplicity, and assume that quantities such as $r$, $\rho$, etc. are dimensionless. In the main text, we restore factors of $\rm{M_{BH}}$ where applicable.

First, we note that the spike computation as laid out in~\cite{Sadeghian:2013laa,Ferrer:2017xwm} assumes the DM particles to move along Kerr geodesics. In other words, the effect the DM has on the metric is relegated to a higher order effect. We thus consider the background to be dominated by the central Kerr BH by taking $g_{\mu\nu}=g_{\mu\nu}^{\rm K}+\epsilon g_{\mu \nu}^{\rm env}$, with $\epsilon$ a bookkeeping parameter which encodes the ``smallness'' of the environment. Following~\cite{Andersson:2022cax}, and restoring factors of $c$ for order-counting purposes, we take Eq.~\eqref{eq:PN_metric_main_text} as our metric. Eq.~\eqref{eq:PN_metric_main_text} is a generic 1PN metric in a harmonic coordinate system, satisfying the relation $\Box x^\mu_{\rm H}=0$, where the subscript (or superscript) $H$ denotes ``harmonic''. The relation between Boyer-Lindquist and harmonic coordinates $(t_{\rm H},r_{\rm H},\theta_{\rm H},\phi_{\rm H})$ may be explicitly found in Ref.~\cite{Hergt:2007ha}.

The stress-energy tensor is taken to be that of a perfect fluid, with the standard form given in Eq.~\eqref{eq:Tmunu_perfect_fluid_main_text}. Since the DM spike solutions are pressureless dust configurations, we set $P=0$. Further, we may write the total energy in terms of the density and the specific internal energy, $\Pi$, as $\mathcal{E}=\epsilon\rho c^2\left(1+\Pi/c^2 \right)$. Since we assume that the DM particles do not have self-interaction, we also take the specific internal energy $\Pi=0$. With these considerations, we are left with a much simpler expression for the stress-energy tensor,
\begin{equation}
\label{eq:Tmunu_perfect_fluid1}
T^{\rm H}_{\mu \nu}= \epsilon\rho u^{\rm H}_\mu u^{\rm H}_\nu,
\end{equation}
where $u^{\rm H}_\mu$ is the four-velocity of the fluid. We note that $T^{\rm H}_{\mu \nu}\sim\mathcal{O}(\epsilon)$, in line with our assumption that the external DM environment only introduces a small correction to the Kerr background. Next, we consider a different observer such that 
\begin{equation}
\label{eq:3velocityrelations}
u_{\rm H}^t=\gamma c, ~~~ u_{\rm H}^i=\gamma v_{\rm H}^i,
\end{equation}
where $v_{\rm H}^i$ is the 3-velocity of the fluid. Conservation of baryon number implies, in this reference frame (see Section IV.A of~\cite{Andersson:2022cax}), that
\begin{equation}
\label{eq:continuity_eqn}
\pd_{t_{\rm H}}\rho^*+\pd_i (\rho^*v_{\rm H}^i)=0,
\end{equation}
where $\rho^*$ is a rescaled density function $\rho^*=\sqrt{-g_{\rm H}}\gamma \rho$.
With the definitions of the modified metric, alongside the normalization condition $u^{\rm H}_\mu u_{\rm H}^\mu=-c^2$, we may show that
\begin{equation}
\label{eq:gamma_expression}
\gamma \approx 1+\frac{v^2}{c^2}+\frac{U}{c^2} +\mathcal{O}\left(c^{-4}\right),
\end{equation}
and 
\begin{equation}
\label{eq:determinant}
\sqrt{-g_{\rm H}}\approx1+\frac{2U}{c^2}+\mathcal{O}(c^{-4}).
\end{equation}
The potentials in Eq.~\eqref{eq:PN_metric_main_text} satisfy a set of Poisson equations which depend on source terms proportional to the rescaled density $\rho^*$, explicitly written in Eq.~\eqref{eq:PN_Poisson_eqs_main_text}. Recalling that the metric is dominated by the Kerr background so that $g_{\mu \nu}=g_{\mu \nu}^{\rm K}+\epsilon g_{\mu \nu}^{\rm env}$, we take $U=U^{\rm K}+\epsilon U^{\rm env}$, with similar definitions for $U_i$ and $\psi$. Expanding Eqs.~\eqref{eq:PN_Poisson_eqs_main_text} in powers of $1/c^2$ and $\epsilon$ yields Poisson equations for the Kerr potentials at $\mathcal{O}(\epsilon^0,1\text{PN})$ sourced by a rotating point mass, and Poisson equations for the environmental potentials at $\mathcal{O}(\epsilon^1,1\text{PN})$ sourced by the external DM. To avoid worrying about the specific form of the Kerr sources, we match $U^{\rm K}, U_i^{\rm K}$, and $\psi^{\rm K}$ to the Kerr metric expanded in harmonic coordinates found in Ref.~\cite{Hergt:2007ha}. This yields, to the order appropriate for Eq.~\eqref{eq:PN_metric_main_text},
\begin{equation}
\label{eq:Kerr_potentials}
\begin{aligned}
U^{\rm K}&=\frac{1}{r_{\rm H}},\\
U_{i}^{\rm K}&=\left\{0,0,\frac{\chi \sin^2\theta_{\rm H}}{2r_{\rm H}}\right\},\\
\psi^{\rm K}&=0.
\end{aligned}
\end{equation}
The $\mathcal{O}(\epsilon^1,1\text{PN})$ equations are listed in the main text in Eqs.~\eqref{eq:PN_Poisson_eqs_main_text1}. To bring the source terms into the form shown there, we replace the fluid velocity terms with the mass currents via the relations $J_\mu=\rho u_\mu$ and Eqs.~\eqref{eq:3velocityrelations}, and rewrite the 3-velocity terms $v^2$ and $v_i$ as 
\begin{equation}
\label{eq:velocity_rewritten}
\begin{aligned}
v_i&=\frac{J_i^{\rm H}}{\rho}\left(1-\frac{v^2}{c^2}-\frac{U^{\rm K}}{c^2} \right)+\mathcal{O}(c^{-4}),\\
v^2&=\delta_{ij}\frac{J^i_{\rm H}J^j_{\rm H}}{\rho^2}\left(1-\frac{2U^{\rm K}}{c^2}- \delta_{kl}\frac{J^k_{\rm H}J^l_{\rm H}}{\rho^2 c^2} \right)+\mathcal{O}(c^{-4}).
\end{aligned}
\end{equation}
The mass currents in harmonic coordinates are related to the Boyer-Lindquist currents given in Eqs.~\eqref{eq:mass_current} via the standard expression $J_\mu^{\rm H}=\frac{\pd x^\nu}{\pd x^\mu_{\rm H}}J_\nu$, where again we use the transformation in Ref.~\cite{Hergt:2007ha}.

We solve Eqs.~\eqref{eq:PN_Poisson_eqs_main_text1} using the standard Green's function approach. For a generic potential $\Phi$ and source term $\mathcal{S}$ obeying the Poisson equation
\begin{equation}
\label{eq:generic_poisson_eq}
\nabla^2\Phi=-4\pi \mathcal{S},
\end{equation}
the solution via direct integration is given by 
\begin{equation}
\label{eq:generic_Poisson_soln}
\Phi(\mathbf{r})=\int \frac{\mathcal{S}(\mathbf{r'})}{|\mathbf{r}-\mathbf{r'}|} \dd^3 x'.
\end{equation}
We first solve Eqs.~\eqref{eq:PN_Poisson_eqs_main_text1} in the harmonic coordinates utilizing Eq.~\eqref{eq:generic_Poisson_soln}, and then swap to Boyer-Lindquist coordinates. Results of this procedure are shown in Fig.~\ref{fig:PN_metric_functions}, where we plot the environmental potentials $U^{\rm env},U_\phi^{\rm env},\psi^{\rm env}$ on the equatorial plane for ``Milky-Way like'' parameters, and several different spins. We note that the potentials exhibit the same scaling with ${\rm{M_{halo}}}, {\rm{r_s}},$ and ${\rm{M_{BH}}}$ as in Eq.~\eqref{eq:spike_fit}. This can be seen by examining the source terms appearing in Eqs.~\eqref{eq:PN_Poisson_eqs_main_text1}, and noting that they are all either proportional to $\rho(r_{\rm H},\theta_{\rm H})$ or $J_{\rm H}(r_{\rm H},\theta_{\rm H})$.

Lastly, we transform the PN metric in harmonic coordinates to the Boyer-Lindquist coordinates via the usual transformation: 
\begin{equation}
g_{\mu \nu}=\frac{\pd x_{{\rm H}}^\alpha}{\pd x^\mu_{\rm BL}}\frac{\pd x_{{\rm H}}^\beta}{\pd x^\nu_{\rm BL}}(g_{\alpha \beta}^{\rm K}+\epsilon g_{\alpha \beta}^{\rm env}).
\end{equation}
To 1PN order, the environmental part of the metric in Boyer-Lindquist coordinates has the following components:
\begin{equation} 
\label{eq:PN_environmental_metric}
\begin{aligned}
g_{tt}^{\rm env,BL}&= \left(1-\frac{2}{r} \right)2U^{\rm env}+\frac{\chi^2\cos\theta \sin \theta}{r^2}\pd_\theta U^{\rm env}\\
&~~~~~~+2\psi^{\rm env}-\left(2-\frac{\chi^2\sin^2\theta}{r} \right)\pd_rU^{\rm env},\\
g_{ti}^{\rm env,BL}&=\left\{0,0,-4U_\phi^{\rm env}\right\},\\
g_{ij}^{\rm env,BL}&=2\delta_{ij}U^{\rm env},
\end{aligned}
\end{equation}
where we have once again suppressed factors of $c$. In line with~\cite{Andersson:2022cax}, for a 1PN metric the $tt$ component is kept at $\mathcal{O}(c^{-4})$, the $ti$ components at $\mathcal{O}(c^{-3})$, and the $ij$ components at $\mathcal{O}(c^{-2})$.

\section{Environmentally modified geodesics}
\label{app:modified_geodesics}

We wish to estimate the effect that the changing background metric, due to the presence of DM spikes, will have on orbiting secondary bodies. To do so, we take our metric to be the Kerr metric modified by a small deviation caused by the environment, $g_{\mu \nu}=g_{\mu \nu}^{\rm K}+\epsilon g_{\mu \nu}^{\rm env}$. We keep the environmental metric generic, but assume that it is axially symmetric and stationary. In line with these assumptions, the only nonzero components of $g_{\mu \nu}^{\rm env}$ are along the components along the diagonal, and $g_{t\phi}$. Further, each of these metric components only depends on the coordinates $(r,\theta)$. The Lagrangian governing the behavior of particles within the Kerr+environment geometry is given by $\mathcal{L}=(1/2)g_{\mu \nu}u^\mu u^\nu$, where $u^\mu=(\dd x^\mu/\dd \tau)$ is the four-velocity of an orbiting particle, and $\tau$ is the proper time. We restrict our analysis to circular equatorial orbits, so we take $\theta=\pi/2$, and set $u^\theta=0$. The Lagrangian becomes, to linear order in $\epsilon$,
\begin{equation}
\begin{aligned}
\label{eq:equatorial_Lagrangian}
2\mathcal{L}&=\left( -1+\frac{2}{r}+\epsilon g_{tt}^{\text{env}}\right)(u^t)^2-2\left(\frac{2 \chi}{r}-\epsilon g_{t\phi}^{\text{env}}\right)u^t u^\phi
\\
&+ \left( \frac{r^2}{r^2-2 r+\chi^2}+\epsilon g_{rr}^{\text{env}}\right)(u^r)^2 
\\
&+\left( r^2+\chi^2+\frac{2\chi^2}{r}+\epsilon g_{\phi \phi}^{\text{env}}\right)(u^\phi)^2\,.
\end{aligned}
\end{equation}
Eq.~\eqref{eq:equatorial_Lagrangian} does not depend on $t$ or $\phi$, and hence it admits two conserved quantities, the energy per unit mass $E$ and $z$-axis angular momentum per unit mass $L_z$, given by 
\begin{equation}
\label{eq:energy_angmom_expressions}
-E=\frac{\partial \mathcal{L}}{\partial u^t},~~~~~L_z=\frac{\partial \mathcal{L}}{\partial u^\phi}.
\end{equation}
Solving Eqs.~\eqref{eq:energy_angmom_expressions} simultaneously for $u^t$ and $u^\phi$, we obtain, to linear order in $\epsilon$, \\ \\
\begin{widetext}
\begin{align}
\label{eq:tdot_phidot}
u^t&= \frac{\chi^2 E (2+r)-2 \chi L_z+E r^3}{r \left(\chi^2+r (r-2)\right)}+ \frac{\epsilon}{r^2 \left(\chi^2+r (r-2)\right)^2} \bigg[ g_{tt}^{\text{env}} \left(\chi^2 (2+r)+r^3\right) \left(\chi^2 E (2+r)-2 \chi L_z+Er^3\right) 
\nonumber
\\
&+g_{t\phi}^{\text{env}} \left(4 \chi^3 E (2+r)+\chi^2 L_z \left(r^2-8\right)+4\chi E r^3+L_z r^3 (r-2)\right)+2 \chi g_{\phi \phi}^{\text{env}} (2 \chi E+L_z (r-2))\bigg]~, 
\\
u^\phi&= \frac{2 \chi E+L_z (r-2)}{\chi^2 r-2 r^2+r^3}+\frac{\epsilon}{r^2 \left(\chi^2+r (r-2)\right)^2} \bigg[ g_{t\phi}^{\text{env}} \left(\chi^2 E \left(8-r^2\right)+4 \chi L_z (r-2)+E r^3 (2-r)\right)
\nonumber
\\
&+2 \chi g_{tt}^{\text{env}} \left(\chi^2 E (2+r)-2 \chi L_z+E r^3\right)+(2
-r) g_{\phi \phi}^{\text{env}} (2 \chi E+L_z (r-2)) \bigg]\,.
\end{align}
With these expressions for $u^t$ and $u^\phi$, we may rewrite the Lagrangian \eqref{eq:equatorial_Lagrangian} in terms of $E$, $L_z$, $u^r$, and $r$. Doing so, and noting that massive particles obey the equation $2\mathcal{L}=-1$, we obtain the following expression: 
\begin{equation}
    \begin{aligned}
    \label{eq:rdot_eq}
  &(u^r)^2\Big\{r^3+\frac{r \epsilon}{\chi^2+r (r-2)}\bigg[g_{rr}^{\text{env}} \left\{\chi^2+r (r-2)\right\}^2+r \big\{-g_{tt}^{\text{env}}
   \left[\chi^2 (2+r)+r^3\right] 
   -4 \chi g_{t \phi}^{\text{env}}+(r-2) g_{\phi \phi}^{\text{env}}\big\}\bigg] \Big\}\\&~~~~= \left(E^2-1\right)\left(r^3+\chi^2r\right)+2 (L_z-\chi
   E)^2-L_z^2 r+2 r^2+\epsilon \Big[g_{tt}^{\text{env}}\left(\chi^2 (2+r)+r \left(L_z^2+r^2\right)\right)
   \\&~~~~~~~~~+2 g_{t \phi}^{\text{env}} (2 \chi+E L_z r)+g_{\phi \phi}^{\text{env}}
   \left(\left(E^2-1\right) r+2\right)\Big]\,.
   \end{aligned}
    \end{equation}
For circular equatorial geodesics, $u^r=0$ and $\dd u^r/\dd \tau=0$. Hence, we set the RHS of Eq.~\eqref{eq:rdot_eq} and its derivative to 0. Doing so allows us to solve for $E$ and $L_z$ as a function of $r$. We obtain, to linear order in $\epsilon$, 
\begin{equation}
    \label{eq:orbital_E_and_Lz}
    \begin{aligned}
    E_{\rm orbit}&=\frac{1-2/r+\chi/r^{3/2}}{\sqrt{1-3/r+2\chi/r^{3/2}}}\bigg[1-\frac{\epsilon}{4 r^{3/2} \left(2 \chi+(r-3) \sqrt{r}\right) \left(\chi+(r-2) \sqrt{r}\right)}\bigg\{\sqrt{r} \left[\chi^2+(r-2) r\right]\bigg\{\left(\chi+r^{3/2}\right)^2 g_{tt}'^{\text{env}}
    \\
    &+2 \left(\chi+r^{3/2}\right) g_{t\phi}'^{\text{env}}+g_{\phi \phi}'^{\text{env}}\bigg\}-2 \left(\chi+r^{3/2}\right) \left[\chi^2-2 \chi \sqrt{r} (r+1)-(r-4) r^2\right] g_{tt}^{\text{env}}
    \\
    &-4\left(\chi^2-2 \chi \sqrt{r}+r^2\right) g_{t\phi}^{\text{env}}-2 \left(\chi+(r-2) \sqrt{r}\right) g_{\phi \phi}^{\text{env}} \bigg\}\bigg]\,,
    \\
   L_{z,\rm orbit}&=\frac{r^{1/2}+\chi^2/r^{3/2}-2\chi/r}{\sqrt{1-3/r+2\chi/r^{3/2}}}\Bigg[1-\frac{\epsilon \sqrt{2 \chi r^{3/2}+(r-3) r^2}}{4 r^{9/4} \left(2 \chi+(r-3) \sqrt{r}\right)^{3/2} \left(\chi^2-2 \chi \sqrt{r}+r^2\right)}\bigg\{ \sqrt{r} \left[\chi^2+(r-2) r\right] \left(\chi+r^{3/2}\right)^3 g_{\text{tt}}'^{\text{env}}
   \\
   &+2 \sqrt{r} \left[\chi^2+(r-2) r\right]
   \left(\chi+r^{3/2}\right)^2 g_{t\phi}'^{\text{env}}+\sqrt{r} \left[\chi^2+(r-2) r\right] \left(\chi+r^{3/2}\right) g_{\phi
   \phi }'^{\text{env}}
   \\
   &-2 \left[\chi^2+2 \chi \sqrt{r} (2 r-1)+r^2 (2 r-5)\right] g_{\phi \phi }^{\text{env}}-2 \left(\chi^2-2 \chi
   \sqrt{r}+r^2\right) \left(\chi+r^{3/2}\right)^2 g_{tt}^{\text{env}}
   \\
   &-4 \left[\chi+(r-2) \sqrt{r}\right] \left(\chi+r^{3/2}\right)^2 g_{t\phi
   }^{\text{env}}\bigg\} \Bigg]\,.
   \end{aligned}
\end{equation}
Finally, with expressions in hand for the energy, angular momentum, $u^t(E,L_z)$, and $u^\phi(E,L_z)$, we compute the orbital frequency of a particle in our Kerr+environmental geometry as a function of the orbital radius. Taking the angular velocity as $\Omega=(d\phi/dt)=u^\phi/u^t$, we obtain the simple relation (to linear order in $\epsilon$):
\begin{equation}
\label{eq:modified_orbital_frequency}
\Omega=\Omega_0-\epsilon \frac{\sqrt{r}}{4} \left[g_{tt}'^{\text{env}}(r)+2\Omega_0 g_{t \phi}'^{\text{env}}(r)+\Omega_0^2 g_{\phi \phi}'^{\text{env}}(r)\right]\,, 
\end{equation}
with $\Omega_0=1/(r^{3/2}+\chi)$. The prime denotes differentiation w.r.t. $r$, and the metric functions are evaluated on the equatorial plane. 

\end{widetext}

\bibliography{refs}
\end{document}